\documentclass[journal,draftclsnofoot,onecolumn,12pt]{IEEEtran}
\usepackage{amsthm,amssymb,graphicx,multirow,amsmath,color,amsfonts}
\usepackage{setspace}	
\usepackage[update,prepend]{epstopdf}
\usepackage[noadjust]{cite}
\usepackage[latin1]{inputenc}
\usepackage{tikz}
\usepackage{pdfpages}
\usepackage{subfigure,array}
\usepackage{tabulary}
\usepackage{multirow}
\usepackage{float}
\usepackage{comment}

\usepackage{mathtools}
\let\subparagraph\relax
\usepackage[compact]{titlesec}
\titlespacing{\section}{0pt}{*1}{*1}
\titlespacing{\subsection}{0pt}{*1}{*1}

\include{notation}
\allowdisplaybreaks 
\allowdisplaybreaks

\def\l{{\ell}}
\def\n{{\rm n}}

\def\nb0{{\mathbf{0}}}
\def\nb1{{\mathbf{1}}}

\def\nbbE{{\mathbb{E}}}

\def\nbbP{{\mathbb{P}}}

\newtheorem{lemma}{Lemma}

\newtheorem{ndef}{Definition}

\newtheorem{theorem}{Theorem}
\newtheorem{prop}{Proposition}
\newtheorem{cor}{Corollary}

\newtheorem{remark}{Remark}
\newtheorem{assumption}{Assumption}
	
\def\P{\mathbb{P}}

\def\pr{\mathtt{P_r}}
\def\p{p}

\def\R{\mathbb{R}}

\def\sinr{\mathtt{SINR}}			
\def\snr{\mathtt{SNR}}

\graphicspath{{./Fig/}}
\title{Millimeter
Wave Integrated Access and Backhaul  in 5G: Performance Analysis and Design Insights}
\author{
Chiranjib Saha and Harpreet S. Dhillon 
\thanks{The authors are with Wireless@VT, Department of ECE, Virginia Tech, Blacksburg, VA, USA. Email: \{csaha,   hdhillon\}@vt.edu. The support of the US National Science Foundation (Grant CNS-1617896) is gratefully acknowledged. 
} }

\begin{document}
\maketitle
\begin{abstract}
With the emergence of integrated access and backhaul (IAB) in the fifth generation (5G) of cellular networks, backhaul is no longer just a passive capacity constraint in cellular network design. In fact, this tight integration of access and backhaul is one of the key ways in which 5G millimeter wave (mm-wave) heterogeneous cellular networks (HetNets) differ from traditional settings where the backhaul network was designed independently  from the radio access network (RAN).  With the goal of elucidating key design trends for this new paradigm, we develop an analytical framework for a two-tier HetNet with IAB where the macro base stations (MBSs) provide  mm-wave backhaul  to the small cell base stations (SBSs). For this network, we derive the downlink rate coverage probability for two types of resource allocations at the MBS:  1) {\em integrated resource allocation (IRA):} where the total bandwidth (BW) is dynamically split between access and backhaul, and 2) {\em orthogonal resource allocation (ORA):} where a static partition is defined for the access and backhaul communications. Our analysis concretely demonstrates that offloading users from the MBSs to SBSs may not provide similar rate improvements in an IAB setting as it would in a HetNet with fiber-backhauled SBS. 
 Our analysis also shows that it is not possible to improve the user rate in an IAB setting by simply densifying the SBSs due to the bottleneck on the rate of wireless backhaul links between MBS and SBS. 
\end{abstract}
\section{Introduction}
Aggressive frequency reuse achieved through network densification is regarded as one of the most effective ways of increasing network capacity. The introduction of low power SBSs has made it possible, in principle, to implement this at large scale in cellular networks. Despite all the promising gains, the number of SBSs actually deployed in practice has lagged the market estimates~\cite{thompson20145g}. This is a direct consequence of the challenges involved in providing reliable backhaul to tens of thousands of these SBSs. 
While it is not viable to connect all the  SBSs to the network core with the traditional fiber backhaul, the wireless backhaul solutions have  not also been widely adopted due to the spectrum shortage at sub-6 GHz. However, thanks to the availability of huge spectrum in mm-wave, it is possible to achieve fiber-like performance on the MBS-SBS backhaul links  {while} keeping sufficient bandwidth for the base station (BS)-user access links. Further, the access and backhaul networks  can be tightly integrated to manage the dynamic traffic demand of the HetNet by proper resource partitioning within the access and backhaul links~\cite{SahaBackhaul,saha2017integrated,dhillon2012modeling,OffloadingSingh}. 
This IAB architecture introduces several new modeling aspects which were not present in the conventional HetNet models with no backhaul constraints on the SBSs. 
 For instance, 
the end-user data rate is affected by the rate achievable on  the wireless backhaul links and  the  number of users and SBSs sharing the  available BW which is significantly different from the number of users served by the BSs  (also known as the {\em load} on the BSs) in the conventional networks. In this paper, we capture these unique  IAB characteristics by designing   the first stochastic geometry (SG)-based multi-cell framework for a two-tier IAB-enabled mm-wave network where the MBSs serve the users and SBSs from the same pool of spectral resources. Using this framework, we seek the answers to the following questions: (i) Should the resources be split between access and backhaul {\em a priori} or allocated dynamically based on the load? (ii) How do the data rates change with network densification under backhaul constraints imposed by IAB? and (iii) How
 effective is offloading traffic from MBSs to SBSs in an IAB setting? 
 \subsection{Background and related works}
Due to the availability of huge bandwidth and the use of noise-limited directional transmission in mm-wave, 5G is envisioning the integration of mm-wave  wireless backhaul network and  RAN such that the same spectral resources and infrastructure could be used for both~\cite{accessbackhaul3gpp}. 
%
   This emerging IAB architecture has motivated a lot of recent research activities, such as  finding optimal routing and scheduling strategies in a mm-wave IAB network \cite{polese2018distributed, m2018maxmin,Delayaware}, end-to-end network simulator design~\cite{polese2018end}, and finding optimal user association schemes for HetNets with IAB~\cite{DeRenzoBackhaul}. 
 The existing works on IAB mostly ignore the effect of 
network topology and its interplay with user traffic, which collectively have a significant  impact on   the signal-to-interference-and-noise-ratios ($\sinr$s) of access and backhaul links as well as the loads on  different BSs. Such spatial interactions can be naturally  captured by SG-based models~\cite{dhillon2012modeling}, where the BS and user locations are modeled as point processes, most commonly the Poisson point processes (PPPs).  These models have yielded tractable expressions of network performance metrics such as coverage~\cite{Madhusudhanan2014,DiRenzoUplink2016}, cell load~\cite{Singh_association_cell} and rate~\cite{OffloadingSingh,Rate6658810} for sub-6 GHz networks. 
However, most of the prior works in this direction focus on the access network performance without incorporating any backhaul capacity constraints. Some notable works that do include backhaul constraints are \cite{suryaprakash2014analysis,SinghResourcePartition,DiRenzoBackhaulHyperdense,Ganti-self-backhaul,tabassum2016analysis,QuekBackhaul,DhillonCaire2015}, where \cite{suryaprakash2014analysis,SinghResourcePartition,DiRenzoBackhaulHyperdense,Ganti-self-backhaul,tabassum2016analysis,DhillonCaire2015} characterize the  network performance in terms of data rate and \cite{QuekBackhaul} in terms of delay. 

 
These SG-based models, initially  applied to sub-6 GHz networks, have been extended  to  the coverage analysis for mm-wave networks~\cite{Bai_mmWave,direnzo2015stochastic,mmWaveHetNetTurgut}. However, none of these works consider the impact of  limited backhaul capacity (of which IAB is a special case). In fact, the SG-based models for mm-wave IAB are quite sparse  with~\cite{SahaBackhaul,saha2017integrated,SinghKulkarniSelfBackhaul} being the only notable related works.  While  \cite{SahaBackhaul,saha2017integrated}  focused on a single macro cell of a two-tier mm-wave IAB,  \cite{SinghKulkarniSelfBackhaul} presents a SG-based multi-cell model of a {\em single-tier}   
mm-wave IAB,  where the BSs and
users are distributed as PPPs. As will be evident in the sequel, none  of these models is sufficient to analyze the rate performance of IAB in a mm-wave mutli-cell {\em multi-tier} network.  In particular,
the aspects of load balancing, which is one of the key flexibilities of HetNets \cite{OffloadingSingh},   has never 
been studied in a multi-cell IAB-enabled HetNet setting.

{Before we state our main contributions, it would be instructive to discuss the fundamental challenges involved in  developing an analytical framework for {data rate in}  mm-wave IAB-enabled HetNets.  
For characterizing  data  rate,  
 one needs to take into account  the $\sinr$ and the cell loads.  
Now the load modeling requires the characterization of the  association cells of the BSs which in mm-waves are fundamentally different  from the relatively well-understood association cells in the sub-6 GHz due to the sensitivity of mm-wave propagation to  blockages (as will be illustrated in Fig.~\ref{fig::association::cell::correlated::blocking}).}  
 The existing approach for  blockage modeling is to assume  that  each link  undergoes independent blocking. This simple assumption turns out to be reasonably accurate (especially when the blockages are not too big) for the characterization of $\sinr$ distribution of a typical receiver, or the coverage probability~\cite{AndrewsMMWave}. Since this assumption facilitates analytical tractability, the follow-up works on mm-wave networks,  including the prior arts on  rate analyses in mm-wave networks~\cite{Kulkarni_backhaul_asilomar,Elshaer_mmwave_association,SinghKulkarniSelfBackhaul}, tend to simply accept the independent blocking  as a {\em de facto} model for blocking.      
However, this assumption may not lead to a meaningful characterization of mm-wave association cells (as will be illustrated in Fig.~\ref{fig::association::cell}) and hence the load served by different BSs. For instance, by ignoring this correlation, two adjoining points in space may be assigned to the cells of two different BSs, thus resulting in association cells that deviate significantly from reality. Therefore, for the association cells, we need to jointly consider the blocking statistics for adjacent points which is likely to have some spatial correlation. While this spatial correlation can be introduced by considering some spatial distribution of blockages~\cite{Bai_mmWave}, it induces tremendous complexity in computing the link state between any transmitter and receiver (line-of-sight (LOS) or non-LOS (NLOS)) and is neglected in all analytical and
even most of the 3GPP simulation models~\cite{3GPPNR}.   
  Therefore, a tractable and reasonably accurate approach to rate analysis needs to revisit such assumptions  for different components of the analysis, while making sure that the resulting constructs remain physically meaningful. Constructing such an approach is the main focus of this paper.  
  
\subsection{Contributions and outcomes}\label{subsec::contributions}
{\em 1) Tractable model for IAB-enabled mm-wave HetNet:}   
We develop a tractable and realistic model for analytically characterizing the performance of an IAN-enabled HetNet operating in the mm-wave frequencies. We assume that only the MBSs have access to the fiber backhaul while the SBSs are wirelessly backhauled by the MBSs over mm-wave links. For this IAB setting, we derive the rate coverage, or equivalently, the CCDF of the downlink data rate perceived by a user equipment (UE) for
two resource partition strategies at the MBS: (a) IRA: where the total BW is dynamically split between access and backhaul, and (b) ORA: where a static partition is defined for the access and backhaul communications. While inclusion of correlated blocking is known to be intractable in the mathematical analyses and unscalable in system-level simulations, we propose a novel way to incorporate its effects into the formulation of cell load, which is an integral part of the rate characterization. Another key novelty of our analysis is the characterization of the joint distribution of the SINRs of the access and backhaul link when the typical UE associates to the SBS. Using these results, we finally derive tractable expressions of rate coverage for both IRA and ORA.

{\em 2) System design insights:} 
Our results provide he following key system design insights. 

(i) As expected, the BW split between access and backhaul links has a  significant impact on the performance of ORA. Our numerical results indicate that there exists an optimum BW split for which the rate coverage is maximized. As SBS density increases, the optimal split claims more BW to be dedicated to the backhaul links. 

(ii) The two-tier IAB network performs better than the single-tier macro-only network but significantly worse than a two-tier network with fiber-backhauled SBSs. Moreover,   offloading users from MBSs to SBSs does not yield significant rate improvement as  observed in a two-tier HetNet with fiber-backhauled SBSs. This is because  the UEs offloaded to SBSs are actually coming back to the MBS  through the increased backhaul load due to self-backhauling. 
 
(iii) While the rate coverages and median rates improve steadily with SBS density for a two-tier HetNet with fiber-backhauled SBSs, these metrics quickly  saturate with increasing SBS density for an IAB setting because of the capacity bottleneck of the wireless backhaul links. This result indicates that the capacity gains of HetNets are significantly overestimated if no constraint on the SBS backhaul is considered. 
\section{System Model}\label{sec::sys::mod}
\subsection{BS and user locations}\label{subsec::BSueloc}
We consider a two-tier HetNet where the MBSs and SBSs are distributed in
\(\mathbb{R}^2\) according to independent homogeneous PPPs
\(\Phi_{\rm m}\) and \(\Phi_{\rm s}\) with densities \(\lambda_{\rm m}\)
and \(\lambda_{\rm s}\), respectively. All BSs are assumed to operate in
mm-wave regime. The UEs are assumed to be distributed according to a homogeneous PPP $\Phi_{\rm u}$  with intensity  $\lambda_{\rm u}$.

The analysis is done for a \emph{typical} UE which is sampled from
\(\Phi_{\rm u}\) uniformly at random. We shift the origin of our
coordinate system to the location of the typical user. The BS that
serves this user is known as the \emph{tagged BS}.
We assume that the MBSs are equipped with high capacity wired backhaul, i.e., they are connected to the  network core by high speed fibers. On the other hand, the SBSs are wirelessly backhauled by the MBSs over mm-wave links. All BSs operate in open access, i.e., a UE may either connect to an MBS or an SBS depending on the $\max$ power-based association strategy (details in Section~\ref{subsec::cell::association}). 
Thus the UEs are served by one-hop links if they are connected to the MBS and two-hop links if they are connected to the SBS.   We refer to the link between a user and BS as an {\em access link} and to the link between an MBS and SBS as a {\em backhaul link}. 

{\bf Notation.} We will denote a point process and its associated counting measure by the same notation. Thus, if $\Phi$ denotes a point process, then $\Phi(A)$ denotes the number of points of $\Phi$
 falling in $A\in{\cal B}(\R^2)$, where ${\cal B}(\R^2)$ denotes the Borel-$\sigma$ algebra in $\R^2$. Also 
 $|\cdot|$ denotes the Lebesgue measure
in $\R^2$ 
(i.e., for a set $B\subseteq\R^2$,  $|B|$ denotes the area of $B$). 
\subsection{Propagation Model}\label{subsec::prop::assumption}
\subsubsection{{Blockage Model}}\label{subsec::blockage}
\begin{figure*}
\minipage{0.4\textwidth}
  \includegraphics[scale=0.5]{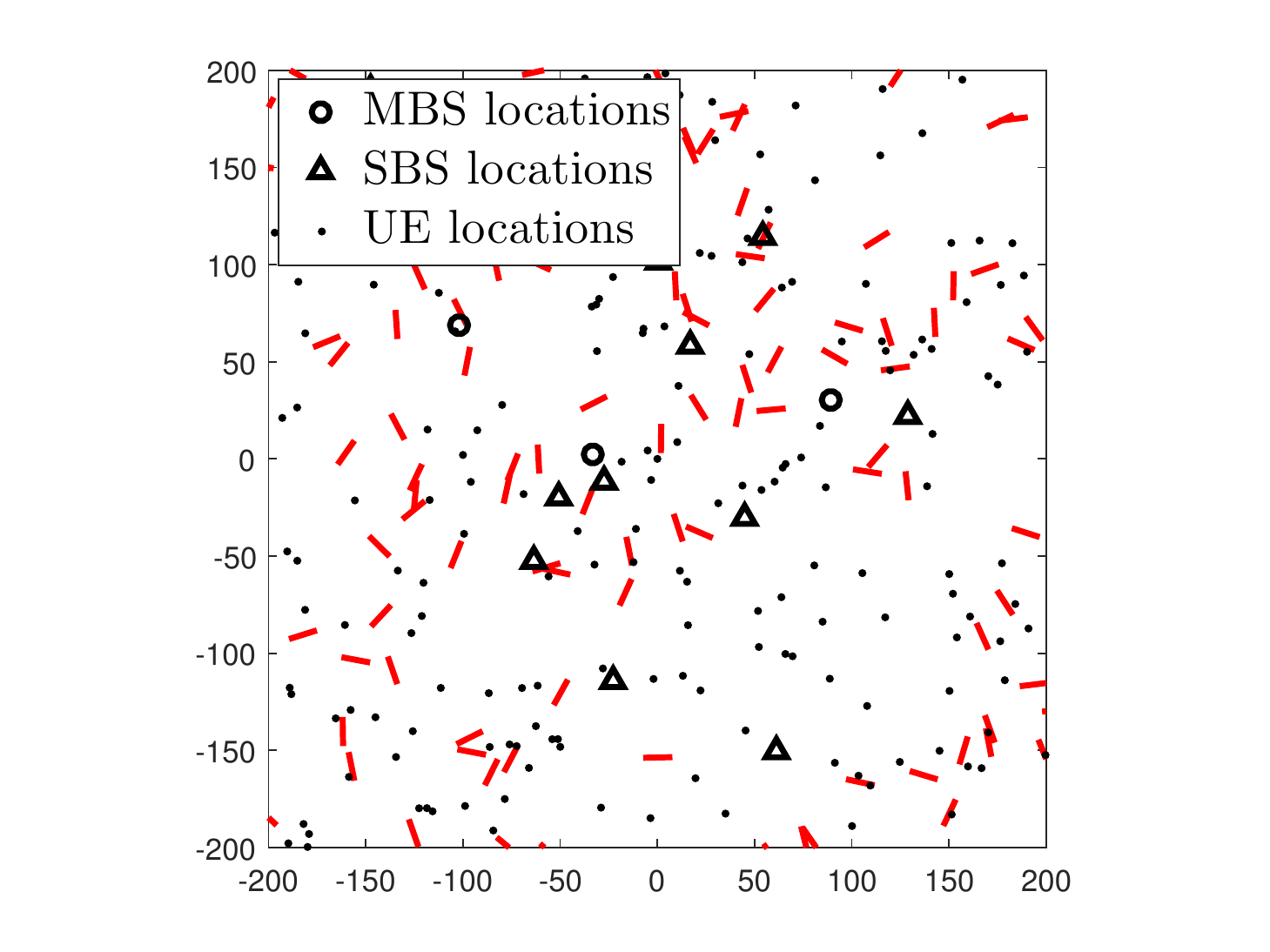}
  \caption{A realization of the two-tier network. The blockages are indicated by red lines.\newline}\label{fig::spatialsetup}
\endminipage\hspace{15pt}
\minipage{0.6\textwidth}
\subfigure[Association to SBS.]{
       \includegraphics[width=.4\linewidth]{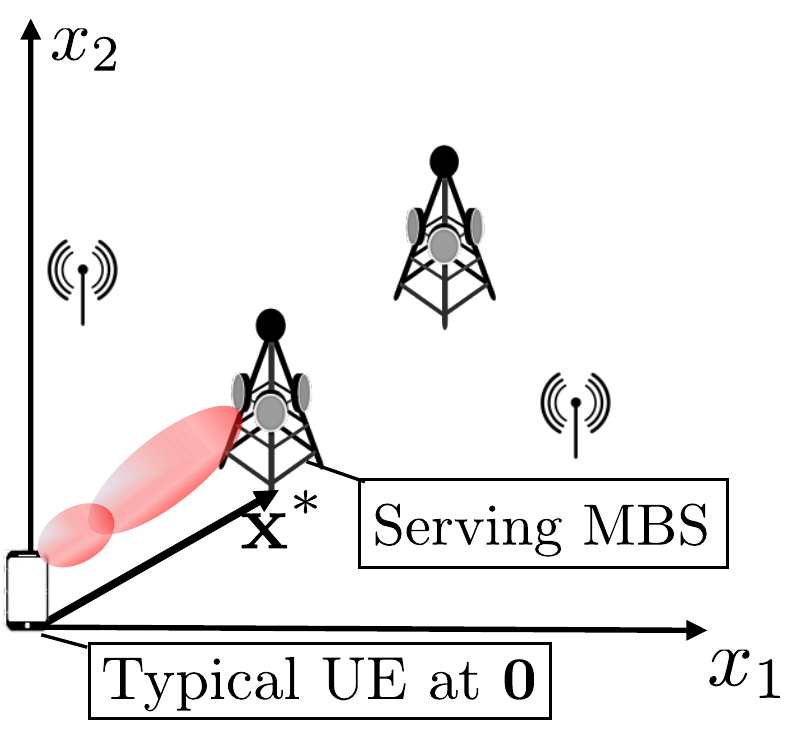}
            \label{fig::association::macro}
        }
\subfigure[Association cells to SBS.]{
       \includegraphics[width=.4\linewidth]{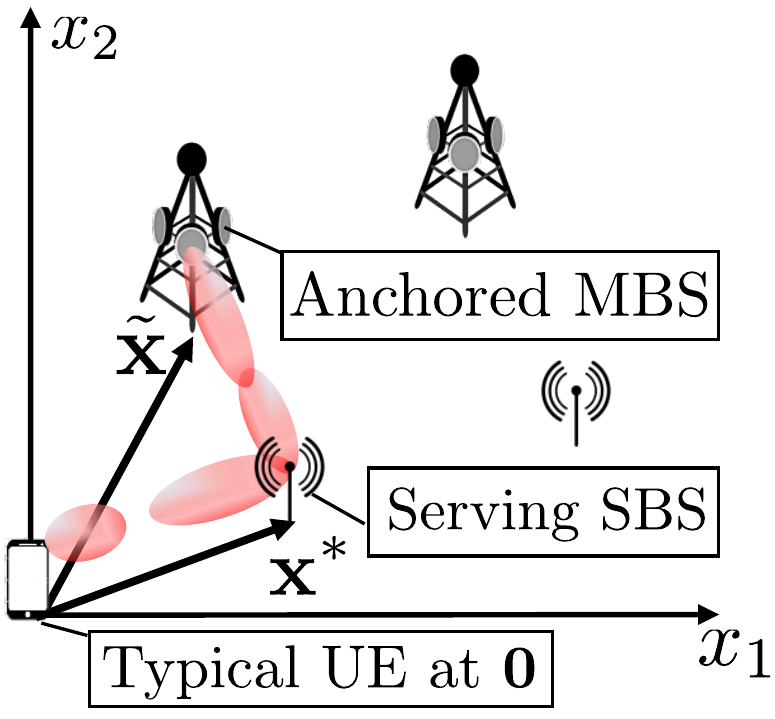}
            \label{fig::association::cell::small}
        }      
 \caption{Illustration of ${\bf x}^*$ and $\tilde{{\bf x}}$.}
             \label{fig::association}
\endminipage
\end{figure*}

Since mm-wave signals are sensitive to physical blockages such as buildings and trees, the LOS  and NLOS pathloss characteristics of mm-wave signals are significantly different. 
Since blockage models are highly context specific, both deterministic~\cite{3GPPNR} and stochastic models~\cite{Bai_mmWave,Aditya_blocking_letter,Aditya_Blindspot} have been used in the literature. Similar to \cite{Aditya_blocking_letter,Aditya_Blindspot}, we will use a well-known stochastic model known as the germ-grain model for modeling blockages.
 In particular,  the blockages are assumed to be a sequence of  line segments  $\Phi_{\rm bl}=\{{\bf p},L_{\rm bl},\theta\}$ where ${\bf p}$, $L_{\rm bl}$, and $\theta$ denote the location of midpoint, length, and orientation of each segment, respectively.  The sequence  $\{{\bf p}\}$ is distributed as a PPP  density $\lambda_{\rm bl}$ in $\R^2$ and $\{\theta\}$ is  a sequence of  independently and identically distributed (i.i.d.) uniform random variables in $(0,2\pi]$.  
 A link between a transmitter at ${\bf x}$  and a receiver at ${\bf y}$ is in LOS ($s({\bf x},{\bf y})= {\l}$) if there is no intersection between the line segment connecting ${\bf x}$ and ${\bf y}$, denoted as $\overline{{\bf x},{\bf y}}$,  and the elements in $\Phi_{\rm bl}$.  We denote the state of a link as $s\in\{{\l},{\rm n}\}$ in accordance with the link being in LOS or NLOS state. 
For  a link of type $k$, the pathloss is defined as
\begin{align} \label{eq::pathloss::germ::grain}
 L_{{k}_i}(z) = 
 \begin{cases}
   z^{\alpha_{{{k}_{i,\l}}}},&\text{if }s=\l,\ i.e.\ \#(\Phi_{\rm bl}\cap\overline{{\bf x},{\bf y}}) = 0, \\
   z^{\alpha_{{{k}_{i,\n}}}},&\text{if }s=\n,\ i.e.\ \#(\Phi_{\rm bl}\cap \overline{{\bf x},{\bf y}}) > 0,
 \end{cases}k\in\{{\rm a},{\rm b}\}, \ i=\{{\rm m},{\rm s}\},
  \end{align}
  where $\{\alpha_{{k}_{i,s}}\}$ denote
the  pathloss exponents and $\#(\Phi_{\rm bl}\cap\overline{{\bf x},{\bf y}})$ gives the number of line segments from $\Phi_{\rm bl}$ that intersect with $\overline{{\bf x},{\bf y}}$. In Fig.~\ref{fig::spatialsetup}, we illustrate a realization of the network. 
 In addition to its relevance from the systems perspective (as justified in \cite{Aditya_Blindspot}), there  
  are two reasons for choosing this particular blocking distribution. First, it reduces to the well-known independent exponential blocking if $L_{\rm bl}$ is not large enough~\cite{Aditya_Blindspot}, which will be useful for the $\sinr$ analysis in Section~\ref{subsec::sinr::distribution}. Second, it is a stationary distribution which will  facilitate the characterization of cell load in Section~\ref{subsec::load::distribution}. 
\subsubsection{Effective Antenna Gain}
The propagation loss in mm-wave frequencies can be overcome by  beamformed directional transmission. To this end,   
 all mm-wave BSs and UEs are assumed to be equipped with antenna arrays. For the analytical tractability, the BS antenna gains are approximated with  sectorized gain patterns, in which the array gains are assumed to be $G_{i}$ for all the angles within the main lobe of beam width $\theta_{b_i}$ and another smaller constant $g_i$ for the rest of the angles ($i\in\{{\rm m},{\rm s}\}$).  The configuration of UE antenna patterns are also assumed to be sectorized with gains $G_{\rm u}$ and $g_{\rm u}$ in the main and side lobes, respectively and beamwidth $\theta_{\rm u}$\footnote{{For notational simplicity, we are  assuming that the antenna units for access and backhaul communications at the SBS and MBS have similar gain patterns. However, different antenna patterns for access and backhaul communications can be easily incorporated without any significant change in the analysis.}}.   Hence, if $\psi_{{k}_i}$ 
  denotes the effective antenna gain for a  link of type $k$ between a BS in $\Phi_{i}$ and a reference point (a  UE for $k={\rm a}$ and SBS for $k={\rm b}$), then under perfect beam alignment between the transmit and receive antennas,   $\psi_{{\rm a}_i}=G_iG_{\rm u}$ and ${\psi}_{{\rm b}_i}= G_{i}G_{\rm s}$. We assume that the BSs are transmitting at constant power spectral density $P_i/W$ ($i\in\{{\rm m},{\rm s}\}$) over the system  BW $W$. 
 Hence, 
the received power over a bandwidth $W'$ in the downlink at a reference point  located at
 ${\bf y}$  from a BS at \({\bf x}\in\Phi_i\) is given by
\begin{equation}
{\rm P}({\bf x},{\bf y})=
\frac{P_{i}}{W}W'h_{{\bf x},{\bf y}} \beta_{{k}_i} \psi_{{k}_{i}} L_{{k}_i}(\|{\bf x}-{\bf y}\|)^{-1}, \qquad  i\in\{{\rm m},{\rm s}\},
\end{equation}
where  $k\in\{{\rm a},{\rm b}\}$ refer to the access and backhaul links, respectively, and  $\beta_{{k}_i}$ is the propagation loss at a reference distance (1 m). We assume each link undergoes Rayleigh fading, i.e., 
  $\{h_{{\bf x},{\bf y}}\}$  is a sequence of i.i.d. random 
  variables  with $h_{{\bf x},{\bf y}}\sim\exp(1)$. 
   Note that while one can, in principle, include more general fading distributions, such as Nakagami~\cite{AndrewsMMWave}, the additional complexity of the analytical expressions will significantly outweigh any additional design insights. This is primarily because the performance {\em trends} are somewhat robust to the choice of fading distribution as long as the distance-dependent channel components are included. The well-known tractability of Rayleigh distribution has therefore led to its use in the analysis of mm-wave systems as well~\cite{SinghKulkarniSelfBackhaul,Elshaer_mmwave_association}.   
\subsection{Association Policy}\label{subsec::cell::association}
The typical UE connects to the BS at ${\bf x}^*$ providing maximum
biased average received power, 
\begin{equation}\label{eq::association::policy}
{\bf x}^* = \arg\max\limits_{\substack{{\bf x}\in\Phi_{i}\\i\in\{{\rm s},{\rm m}\}}}P_i T_i \beta_{{\rm a}_i} G_{i} G_{\rm u} L_{{\rm a}_i}(\|{\bf x}\|)^{-1},
\end{equation}
where $T_i$ denotes the bias factor for association to the $i^{th}$ BS-tier~\cite{OffloadingSingh}. As it will be demonstrated in the sequel, bias factors play pivotal role to offload users (traffic) from one tier to another~\cite{loadbalancingAndrews2014}.  
If the serving BS is an SBS, i.e. ${\bf x}^*\in\Phi_{\rm s}$, then this SBS is wirelessly backhauled to an MBS in $\Phi_{\rm m}$ offering maximum power at the serving SBS location. We call this MBS the {\em anchor MBS} of the serving SBS. Thus,  if $\tilde{\bf x}$ is the location of the anchor  MBS of the SBS at ${\bf x}^*$, then, 
\begin{equation}\label{eq::association::rule::backhaul}
\tilde{\bf x} = \arg\max\limits_{{\bf x}\in\Phi_{\rm m}}P_{\rm m}\beta_{{\rm b}_{\rm m}}G_{\rm m}G_{\rm s}L_{{\rm b}_{\rm m}}(\|{\bf x}-{\bf x}^*\|)^{-1}.
\end{equation} 
Fig.~\ref{fig::association} gives an illustration of ${\bf x}^*$ and $\tilde{\bf x}$.  
Following the association policy, this typical access link will be associated to either an MBS or an SBS with  an association probability which is formally defined as follows.  
\begin{ndef}[Association Probability]\label{def::association::prob}
The association probability ${\cal A}_{i}$ is defined as the probability of the following association event:
$ {\cal A}_{i} = \nbbP\left({\bf x}^*\in \Phi_{i}\right)$, $\forall\ i\in\{{\rm m},{\rm s}\}$.
\end{ndef}
\begin{figure*}
\centering
\subfigure[Association cells for RAN.]{
       \includegraphics[width=.45\linewidth]{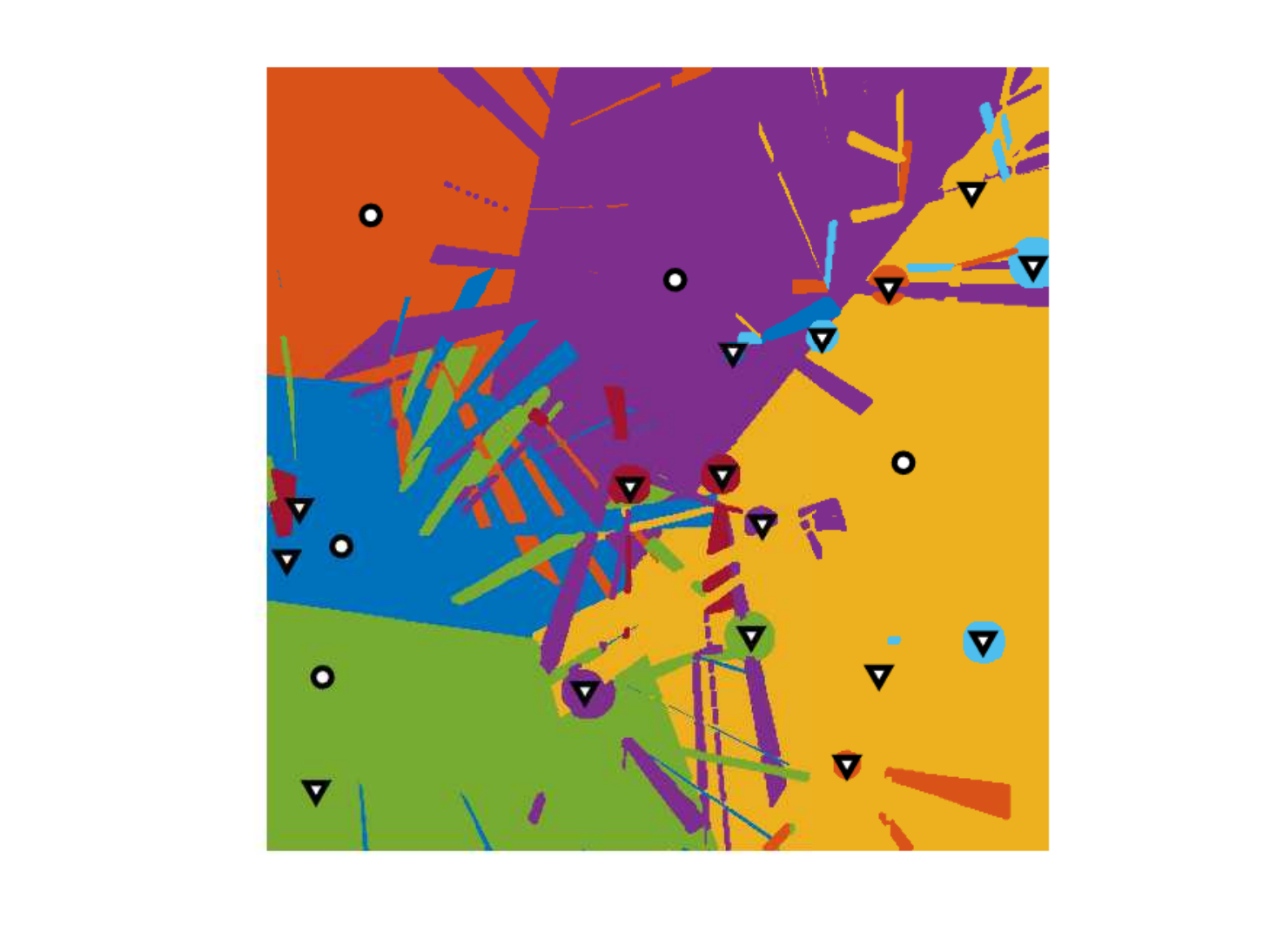}
            \label{fig::association::cell::mmwave}
        }\hfill
\subfigure[Association cells for backhaul network.]{
       \includegraphics[width=.45\linewidth]{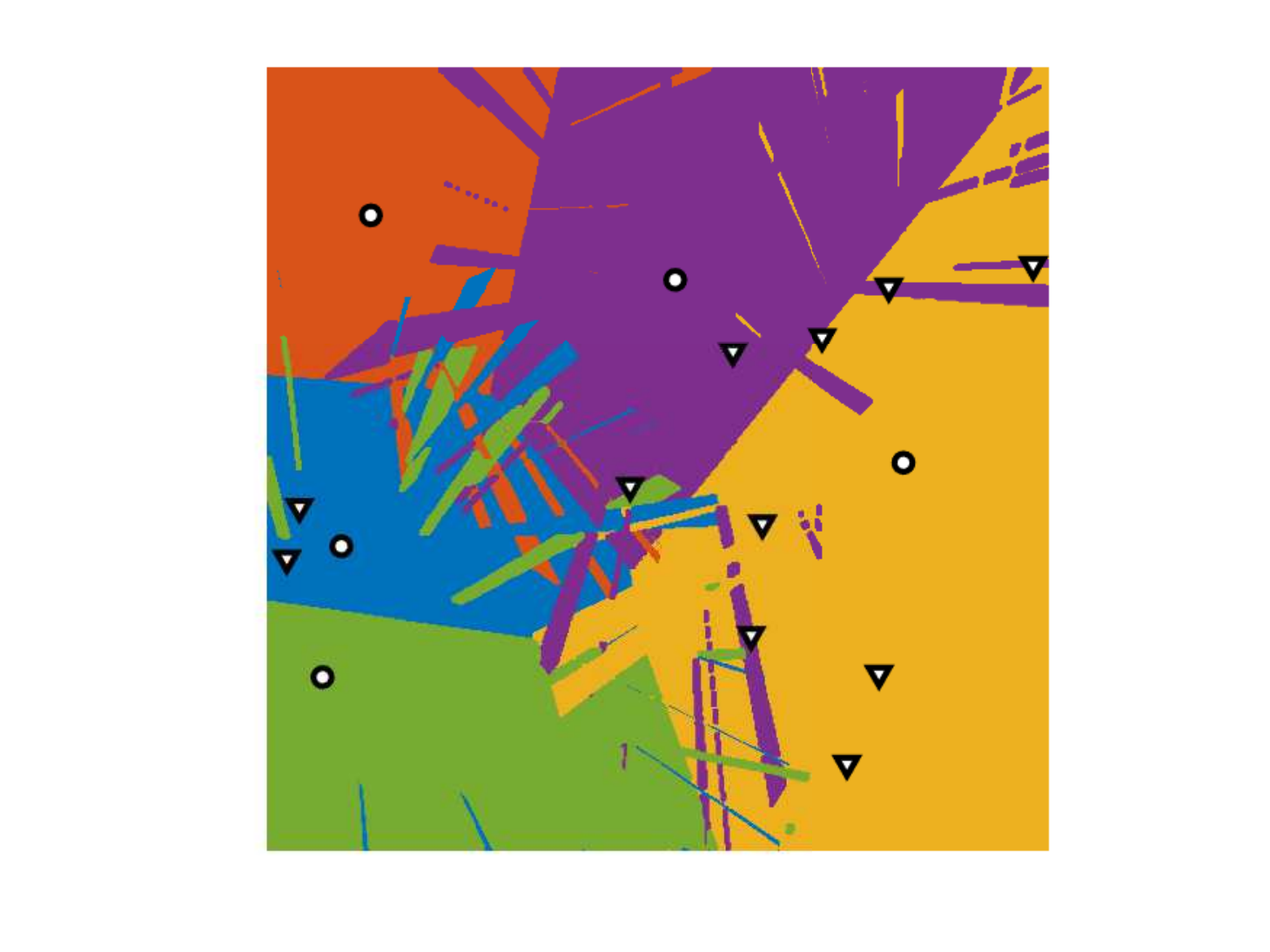}
            \label{fig::association::cell::mmwave::backhaul}
        }      
 \caption{Association cells formed by the BSs of the two-tier HetNet under correlated  blocking. Circles represent the MBSs, and triangles represent the SBSs.}
             \label{fig::association::cell::correlated::blocking}
\end{figure*} 
Since the spatial distribution of the network is stationary,  ${\cal A}_{i}$ denotes the fraction of users of $\Phi_{\rm u}$ being served by the BSs of $\Phi_i$~\cite{OffloadingSingh}. We now define the association cells as follows. 
\begin{ndef}[Association Cell]\label{def::association::cells}
The association cell of a BS located at ${\bf x}$ refers to a closed subset in $\R^2$ where the received power from ${\bf x}$ is greater than the received powers from all other BSs in the network. For the access links, an association cell of the $i^{th}$ tier can be expressed as
\begin{multline}\label{eq::associaiton::cell::access}
{\cal C}_{{\rm a}_i} ({\bf x})= \{{\bf z}\in\R^2:P_iT_i\beta_{{\rm a}_i}G_{{i}}L_{{\rm a}_i}(\|{\bf z}-{\bf x}\|)^{-1}\geq P_jT_j\beta_{{\rm  a}_j}G_{{j}}L_{{\rm a}_j}(\|{\bf z}-{\bf y}\|)^{-1},\\\forall\ {\bf y}\in\Phi_j, j\in\{{\rm m},{\rm s}\}|{\bf x}\in\Phi_{i}\},
\end{multline}
and for backhaul links, 
\begin{equation}\label{eq::association::cell::backhaul}
{\cal C}_{\rm b}({\bf x}) = \{{\bf z}\in\R^2:L_{{\rm b}_{\rm m}}(\|{\bf z}-{\bf x}\|)^{-1}\geq L_{{\rm b}_{\rm m}}(\|{\bf z}-{\bf y}\|)^{-1},\forall\ {\bf y}\in\Phi_{\rm m}|{\bf x}\in\Phi_{\rm m}\}.
\end{equation} 
\end{ndef}
These association cells are depicted in Fig.~\ref{fig::association::cell::correlated::blocking}. 
\subsection{Interference Modeling}
 We now elaborate on the aggregate interference $I_{{\rm a}_{j}}$  from $\Phi_j$ ($j\in\{{\rm m},{\rm s}\}$) experienced by the typical access link, which can be written as
 \begin{align}\label{eq::interference}
 I_{{\rm a}_{j}}= \sum\limits_{{\bf x}\in\Phi_{j}\setminus\{{{\bf x}}^*\}}\frac{P_j}{W}W' h_{0,{\bf x}} \beta_{{\rm a}_j}\psi_{{\rm a}_{j}} L_{{\rm a}_{j}}(\|{\bf x}\|)^{-1} ,
 \end{align}
where $\psi_{{\rm a}_{j}}$ denotes the effective antenna gain of an interfering link seen by the typical UE. 
 Similar to~\cite{Bai_mmWave,mmWaveHetNetTurgut,SinghKulkarniSelfBackhaul}, we model beam directions of the interfering BSs as uniform random variables in $(0,2\pi]$. Then, $\{\psi_{{\rm a}_{j}}\}$ becomes a sequence of i.i.d. discrete random variables taking values from the set ${\cal M}_{{\rm a}_{j}}= \{G_jG_{\rm u}, G_{j}g_{\rm u}, g_jG_{\rm u}, g_jg_{\rm u}\}$ with probabilities $\{\frac{\theta_{b_j}\theta_{b_{\rm u}}}{4\pi^2},\frac{(2\pi-\theta_{b_j})\theta_{b_{\rm u}}}{4\pi^2},\frac{\theta_{b_j}(2\pi-\theta_{b_{\rm u}})}{4\pi^2},\allowbreak\frac{(2\pi-\theta_{b_j})(2\pi-\theta_{b_{\rm u}})}{4\pi^2}\}$, respectively, where $j\in\{{\rm m},{\rm s}\}$. In general, we will denote ${\cal G}\in{\cal M}_{{\rm a}_j}$ as an element  occurring with probability $p_{\cal G}$.
We now shift our attention to the interference experienced by the tagged backhaul link.
Similar to the access links, $\psi_{{\rm b}_j}$ can be modeled as a discrete random variable taking values from the set ${\cal M}_{{\rm b}_{\rm j}} = \{G_jG_{\rm s},g_jG_{\rm s},G_jg_{\rm s},g_jg_{\rm s}\}$ with probabilities $\{\frac{\theta_{b_j}\theta_{b_{\rm s}}}{4\pi^2},\frac{(2\pi-\theta_{b_j})\theta_{b_{\rm s}}}{4\pi^2},\frac{\theta_{b_j}(2\pi-\theta_{b_{\rm s}})}{4\pi^2},\frac{(2\pi-\theta_{b_j})(2\pi-\theta_{b_{\rm s}})}{4\pi^2}\}$, respectively, where $j\in\{{\rm m},{\rm s}\}$.  Using this, the interference experienced by the tagged backhaul link from all BSs in $\Phi_j$ conditioned on ${\bf x}^*\in\Phi_{\rm s}$ can be expressed as
\begin{align}\label{eq::interference::backhaul::macro}
I_{{\rm b}_{ j}}=\begin{cases}
\sum\limits_{{\bf x}\in\Phi_{\rm m}\setminus\{\tilde{\bf x}\}}\frac{P_{\rm m}}{W}W' h_{{\bf x}^*,{\bf x}} \beta_{{\rm b}_{\rm m}}{\psi}_{{\rm b}_{\rm m}}\: L_{{\rm b}_{\rm m}}(\|{\bf x}-{\bf x}^*\|)^{-1},& j={\rm m},\\
\sum\limits_{{\bf x}\in\Phi_{\rm s}\setminus\{{\bf x}^*\}}\frac{P_{\rm s}}{W}W' h_{{\bf x}^*,{\bf x}} \beta_{{\rm b}_{\rm s}}{\psi_{{\rm b}_{\rm s}}}\: L_{{\rm b}_{\rm s}}(\|{\bf x}-{\bf x}^*\|)^{-1},& j={\rm s}.
\end{cases}
\end{align}
We also assume that all BSs are active in the downlink ({\em full buffer} assumption). This means that there is at least one user in an access association cell and one SBS in a backhaul association cell. This assumption is justified since $\lambda_{\rm m}<<\lambda_{\rm s}<<\lambda_{\rm u}$.

\subsection{Resource Allocation and Data Rate}\label{subsec::ResourceAllocation}
In this Section, we introduce two resource allocation strategies. 

\begin{figure*}
\centering
\includegraphics[scale=.57]{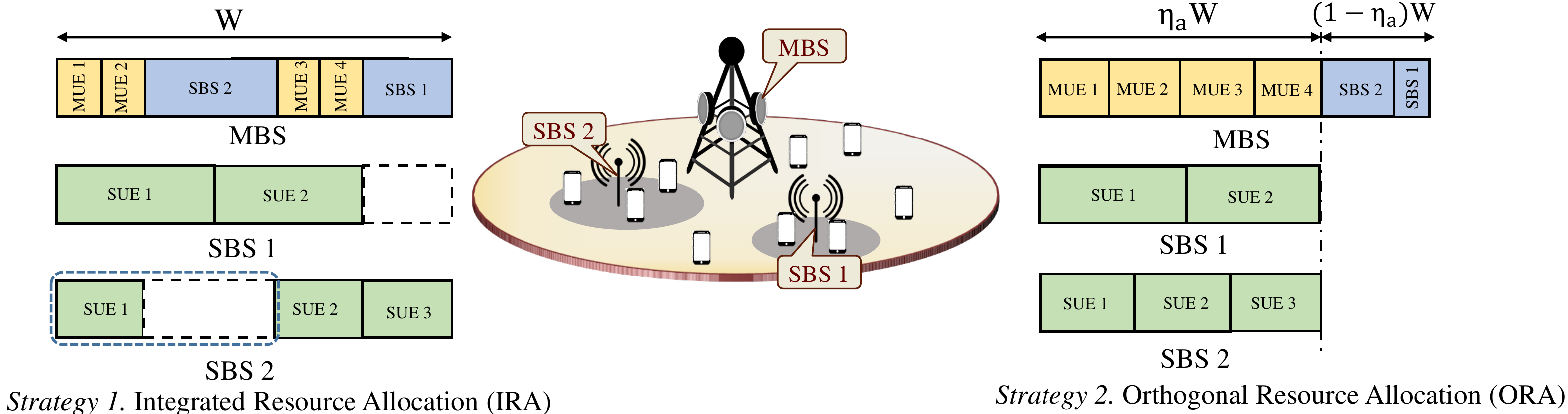}
\caption{Resource partition strategies for a toy example: MBS with two SBSs and four macro users (MUEs), SBS 1 has two users (denoted as SUEs) and SBS 2 has three users.}\label{fig::systemmodel}
\end{figure*}
\subsubsection{Integrated Resource Allocation (IRA)}
We assume that the access and backhaul links share the same pool of radio resources through  orthogonal resource allocation (such as time and frequency division multiple access) and at any BS the total pool of available resources is equally divided among the number of UEs served by each BS (i.e. the BS load) by a simple round robin scheduling.  
 By this  resource allocation scheme, 
if the typical UE connects to the MBS (${\bf x}^*\in\Phi_{\rm m}$), the resource fraction allocated for the typical UE by the tagged MBS is the inverse of the total load on the tagged MBS which is $(\Phi_{\rm u}({\cal C}_{{\rm a}_{\rm m}}({\bf x}^*))+\sum_{\substack{{\bf x}\in\Phi_{\rm s}\cap {\cal C}_{\rm b}({\bf x}^*)}}\Phi_{\rm u}({\cal C}_{{\rm a}_{\rm s}}({\bf x})))$. Here the first term is the load due to the users connected to the tagged MBS over access links and the second terms is due to the users connected to the tagged MBS via SBSs (two hop links). If the typical UE connects to the SBS (${\bf x}^*\in\Phi_{\rm s}$), the fraction of total resources allocated for the tagged SBS for backhaul  by the anchor MBS  is 
\begin{equation}\label{eq::resource::fraction::backhaul::IRA}
\omega = \frac{\Phi_{\rm u}({\cal C}_{{\rm a}_{\rm s}}({\bf x}^*))}{\Phi_{\rm u}({\cal C}_{{\rm a}_{\rm m}}(\widetilde{\bf x}))+\sum\limits_{\substack{{\bf x}\in\Phi_{\rm s}\cap{\cal C}_{\rm b}(\widetilde{\bf x})}}\Phi_{\rm u}({\cal C}_{{\rm a}_{\rm s}}({\bf x}))}.
\end{equation}
 At the tagged SBS,  the resources not occupied by the backhaul link can be further split equally between  the access links of the associated UEs. Thus the rate of a UE is given by 
\begin{align}\label{eq::Rate}
{\tt Rate}_{\rm IRA}  = 
\begin{cases}
 \frac{W}{\Phi_{\rm u}({\cal C}_{{\rm a}_{\rm m}}({\bf x}^*))+\sum\limits_{\substack{{\bf x}\in\Phi_{\rm s}\cap {\cal C}_{\rm b}({\bf x}^*)}}\Phi_{\rm u}({\cal C}_{{\rm a}_{\rm s}}({\bf x}))}\log\left(1+\sinr_{\rm a}({\bf 0})\right),&\text{if }{\bf x}^*\in\Phi_{\rm m},\\
  \frac{W}{\Phi_{\rm u}({\cal C}_{{\rm a}_{\rm s}}({\bf x}^*))}\min\left(\omega\log\left(1+\sinr_{\rm b}({\bf x}^*)\right) ,(1-\omega)\log\left(1+\sinr_{\rm a}({\bf 0})\right)\right),&\text{if }{\bf x}^*\in\Phi_{\rm s}.
\end{cases}
\end{align}
 Here the $\sinr$ on the access link experienced by the typical UE conditioned on the fact that it connects to a BS of $\Phi_i$ (i.e. ${\bf x}^*\in\Phi_i$) is expressed as
\begin{equation}\label{eq::sinr_access}
{\tt SINR}_{\rm a}({\bf 0}) = \frac{P_i G_{i} G_{\rm u} \beta_{{\rm a}_i} h_{0,{\bf x}^*}L_{{\rm a}_i}(\|{\bf x}^*\|)^{-1}}{\sum\limits_{j\in\{{\rm s},{\rm m}\} }\sum\limits_{{\bf x}\in\Phi_{j}\setminus\{{{\bf x}}^*\}}{P_j} h_{0,{\bf x}} \beta_{{\rm a}_j}{\psi}_{{\rm a}_{j}} L_{{\rm a}_{j}}(\|{\bf x}\|)^{-1}+{\tt N}_0W},
\end{equation}
and the $\sinr$ on the backhaul link experienced by the serving SBS conditioned on ${\bf x}^*\in\Phi_{\rm s}$ is expressed as
\begin{align}\label{eq::sinr_backhaul}
{\tt SINR}_{\rm b}({\bf x}^*) = \frac{P_{\rm m} G_{{\rm m}}G_{\rm s} \beta_{{\rm b}_{\rm m}}h_{{\bf x}^*,\tilde{\bf x}}L_{{\rm b}_{\rm m}}(\|{\bf x}^*-\tilde{\bf x}\|)^{-1}}{
\sum\limits_{j\in\{{\rm m},{\rm s}\}}
\sum\limits_{{\bf x}\in\Phi_{\rm j}\setminus\{\tilde{\bf x}\}}
{P_{j}} h_{{\bf x}^*,{\bf x}} \beta_{{\rm b}_{j}}{\psi}_{{\rm b}_{j}} 
L_{{\rm b}_{j}}(\|{\bf x}-{\bf x}^*\|)^{-1}
+{\tt N}_0W}.
\end{align}
\subsubsection{Orthogonal Resource Allocation (ORA)}
In the ORA scheme, we assume that a fraction $\eta_{\rm a}$ of resources is reserved for access links and the rest is allocated to the backhaul links. The share of the total backhaul BW  $(1-\eta_{\rm a})W$ obtained by an SBS at ${\bf x}$ is proportional to its load ($\Phi_{\rm s}({\cal C}_{{\rm a}_{\rm s}}({\bf x}))$). 
Then the rate of a UE is given by
\begin{align}\label{eq::Rate::ORA}
{\tt Rate}_{\rm ORA} = 
\begin{cases}
 \frac{\eta_{\rm a}W}{\Phi_{\rm u}({\cal C}_{{\rm a}_{\rm m}}({\bf x}^*))}\log\left(1+\sinr_{\rm a}({\bf 0})\right), \ \text{if }{\bf x}^*\in\Phi_{\rm m},\\
  \min\bigg(\frac{\eta_{\rm a}W}{\Phi_{\rm u}({\cal C}_{{\rm a}_{\rm s}}({\bf x}^*))}\log\left(1+\sinr_{\rm a}({\bf 0})\right),\frac{(1-\eta_{\rm a})W}{\sum\limits_{{\bf x}\in {\cal C}_{\rm b}(\tilde{\bf x})} \Phi_{\rm u}({\cal C}_{{\rm a}_{\rm s}}({\bf x}))}\log\left(1+\sinr_{\rm b}({\bf x}^*) \right)\bigg),
  \text{if }{\bf x}^*\in\Phi_{\rm s},
\end{cases}
\end{align}  
where $\sinr_{\rm a}({\bf 0})$ is given by \eqref{eq::sinr_access} and $\sinr_{\rm b}({\bf x}^*)$ is given by 
\begin{equation}
{\tt SINR}_{\rm b}({\bf x}^*) = \frac{P_{\rm m} G_{{\rm m}}G_{\rm s} h_{{\bf x}^*,\tilde{\bf x}}\beta_{{\rm b}_{\rm m}}L_{{\rm b}_{\rm m}}(\|{\bf x}^*-\tilde{\bf x}\|)^{-1}}
{\sum\limits_{{\bf x}\in\Phi_{\rm m}\setminus\{\tilde{\bf x}\}}{P_{\rm m}} h_{{\bf x}^*,{\bf x}} \beta_{{\rm b}_{\rm m}}{\psi_{{\rm b}_{\rm m}}}\: L_{{\rm b}_{\rm m}}(\|{\bf x}-{\bf x}^*\|)^{-1}
+{\tt N}_0W}.
\end{equation}
Note that $\sinr_{\rm b}({\bf x}^*)$ for  ORA is greater than $\sinr_{\rm b}({\bf x}^*)$ for  IRA, since in ORA, the tagged backhaul link operating in backhaul BW will not experience the interference from SBSs operating in access BW. However, it will be shown in the sequel that this interference difference does not affect the rate since the backhaul links are mostly noise limited.  Under the {\em full buffer} assumption, $\sinr_{\rm a}({\bf 0})$-s for IRA and ORA are the same.  
We define the rate coverage probability (or simply rate coverage) as the complementary cumulative density function (CCDF) of rate, i.e., $\pr = \nbbP({\tt Rate}>\rho)$, where $\rho$ is the target rate threshold. The two resource allocation strategies are illustrated in Fig.~\ref{fig::systemmodel}.
\subsection{Two-tier HetNet with fiber-backhauled SBSs} To compare and contrast the rate characteristics of the two-tier HetNet with IAB, we define another two-tier network where the SBSs have access to fiber backhaul similar to  the MBSs. This setup is also known as HetNets with {\em ideal SBS backhaul} and has been thoroughly analyzed in the literature~\cite{DownlinkUplinkCellAssociationmmWave,mmWaveHetNetTurgut}. The user perceived rate in this setup can be expressed as:
\begin{align}\label{eq::rate::ideal::backhaul}
{\tt Rate}_{\rm Wb} = 
\begin{cases}
 \frac{W}{\Phi_{\rm u}({\cal C}_{\rm m}({\bf x}^*))}\log\left(1+\sinr_{\rm a}({\bf 0})\right),&\text{if }{\bf x}^*\in\Phi_{\rm m},\\
  \frac{W}{\Phi_{\rm u}({\cal C}_{\rm s}({\bf x}^*))}\log\left(1+\sinr_{\rm a}({\bf 0})\right), &\text{if } {\bf x}^*\in\Phi_{\rm s}, 
\end{cases}.
\end{align}
Clearly it can be seen that ${\tt Rate}_{\rm Wb}$ stochastically dominates ${\tt Rate}_{\rm IRA}$ and ${\tt Rate}_{\rm ORA}$, i.e., $\nbbP({\tt Rate}_{\rm Wb}>\rho)\geq \nbbP({\tt Rate}_{\rm \epsilon}>\rho)$ for $\epsilon \in \{{\rm IRA},{\rm ORA}\}$.  
\section{Rate Distribution}\label{sec::rate::distribution}
In this Section, we evaluate the rate coverage probability defined in the previous Section.  Note that the random variables appearing in the  ${\tt Rate}$ expressions are of  two main types, $\sinr$s of the access and backhaul links  and loads on different BSs. While these $\sinr$ and load variables are correlated  due to the same underlying point processes, this correlation is typically ignored for analytical tractability in this stationary setup without incurring any significant loss in accuray~\cite{Singh_association_cell,Rate6658810,SinghJointRate2015,SinghKulkarniSelfBackhaul}.

\subsection{Load Distribution}\label{subsec::load::distribution}
\begin{figure}
\centering
\subfigure[Association cells for RAN]{
      \includegraphics[width=.45\linewidth]{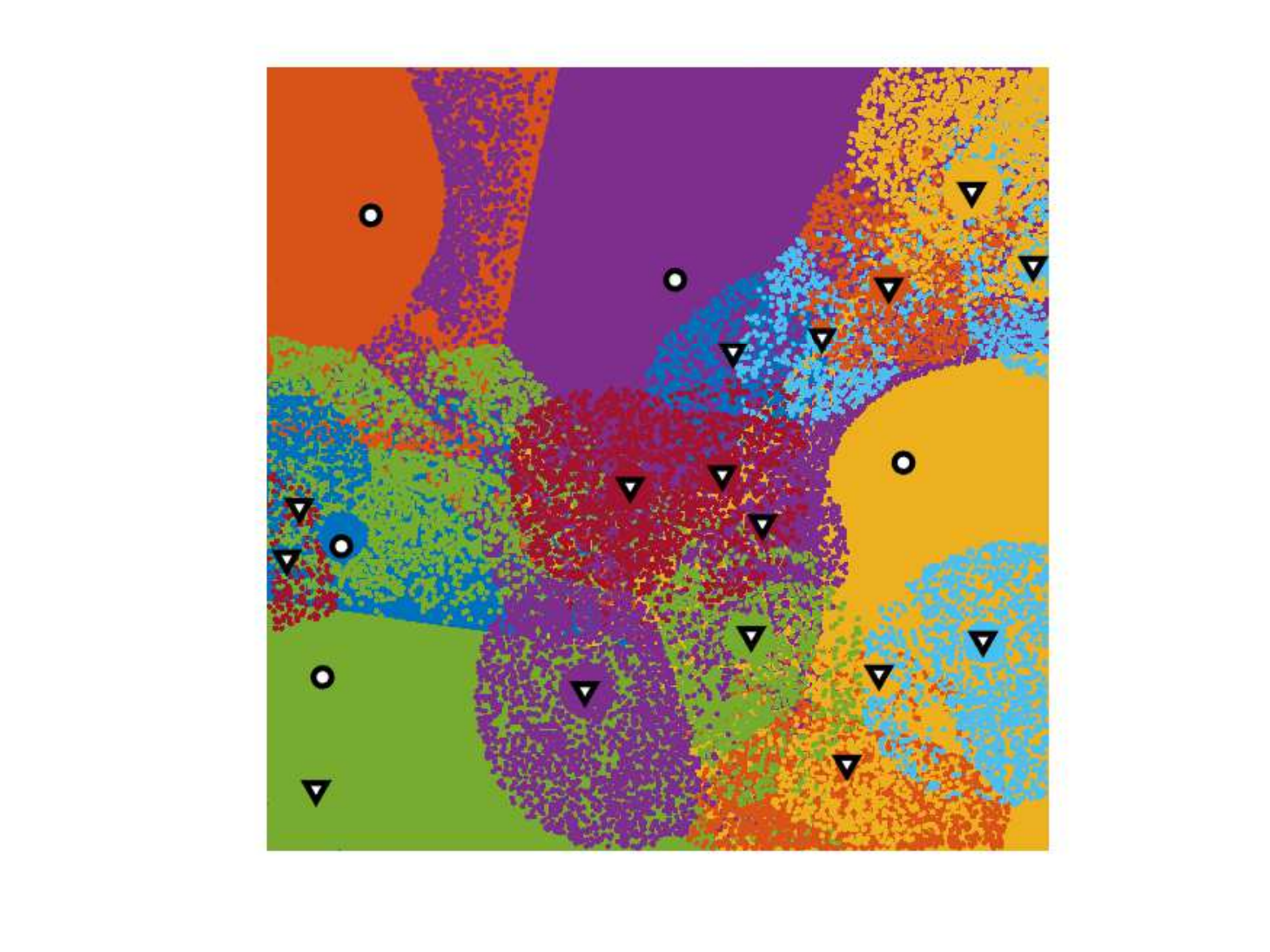}
           \label{fig::association::cell::voronoi}             
       }\hfill
\subfigure[Association cells for backhaul network]{
      \includegraphics[width=.45\linewidth]{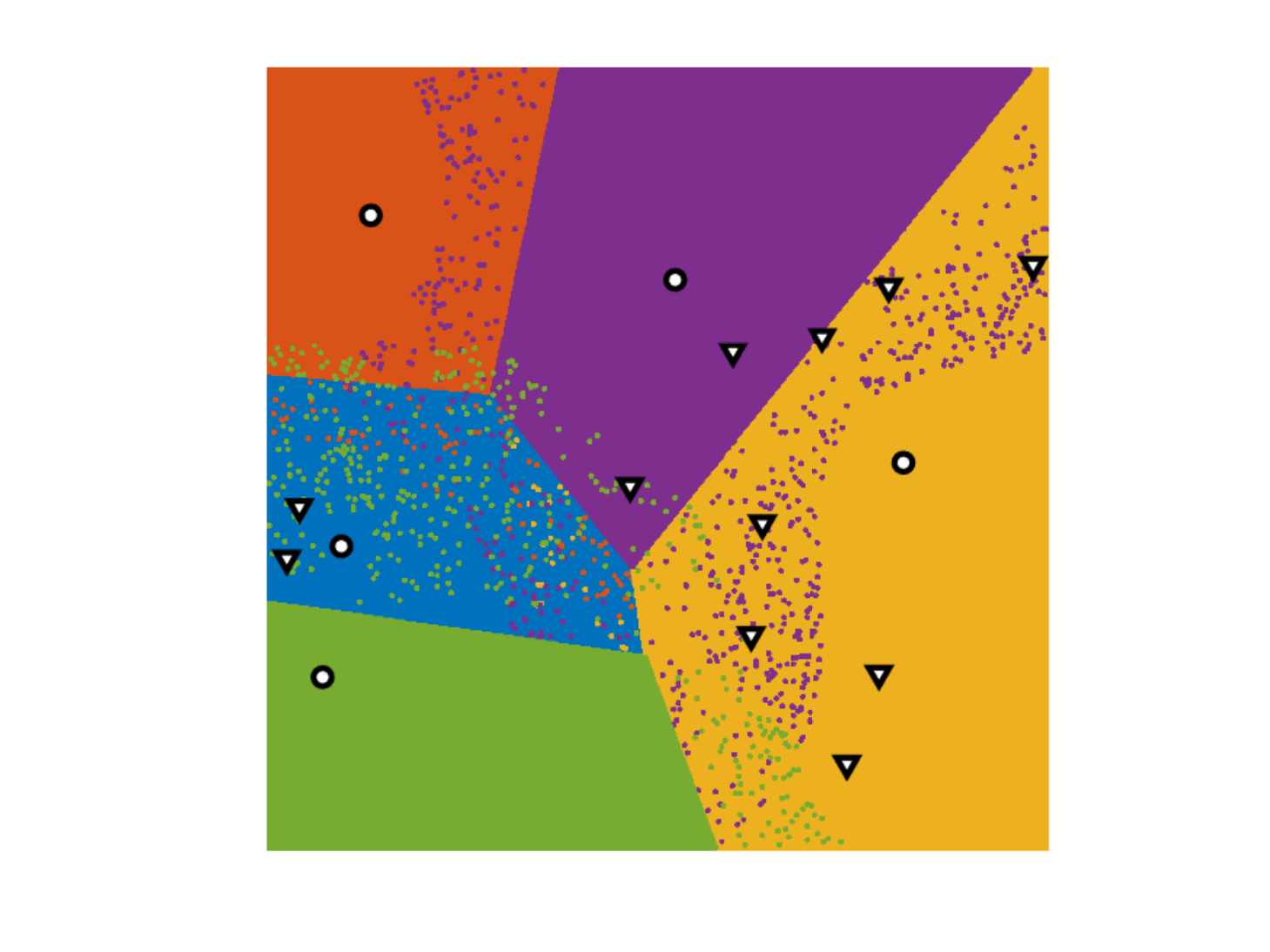}
           \label{fig::association::cell::voronoi::backhaul}             
       }       
 \caption{Association cells formed by the BSs of the two-tier HetNet under  independent blocking. Circles represent the MBSs, and triangles represent the SBSs.}
             \label{fig::association::cell}
\end{figure} 
 We first focus on the distributions of loads on the  serving BS and the anchor MBS which appeared in the expression of rates in \eqref{eq::Rate}, \eqref{eq::Rate::ORA}, and \eqref{eq::rate::ideal::backhaul}.  {As discussed earlier, the independent blocking assumption in mm-wave  is a well-accepted assumption for the $\sinr$ analysis. But it is not quite meaningful for load analysis where it is imperative to consider spatial correlation of the link states to make sure that two adjoining points in space are not assigned to two different association cells.}  
That said, the existing works on  the load characterization in a PPP-network (such as \cite{SinghKulkarniSelfBackhaul}) completely ignore this spatial correlation and simply assume that the link states seen by any two points are completely independent. This assumption leads to the association cells (a key component of the load analysis)  that have no physical significance (such as the ones shown in in Figs.~\ref{fig::association::cell::voronoi} and   \ref{fig::association::cell::voronoi::backhaul}). On the contrary, the association cells have much regular shape (see Figs.~\ref{fig::association::cell::mmwave} and \ref{fig::association::cell::mmwave::backhaul}) if we consider spatial correlation of link states which is induced by the germ-grain model. However,  the exact characterization of the association cells in our current setup is extremely difficult. Note that for a PPP-modeled HetNet in sub-6 GHz, the association cell areas  can be  analyzed  under very simple propagation environment with no blockage-effects which reduces to the formation of  weighted Poisson Voronoi (PV) tessellation~\cite{OffloadingSingh,SinghJointRate2015,Singh_association_cell}. 
While the weighted PV cells may not appear to be  directly applicable to the setting considered in this paper, one can discover some useful connections in order to obtain a tractable characterization of load. The key enabling argument is provided next. 
\begin{remark}\label{rem::stationary}
Since the spatial distribution of blockages is stationary~\cite{Aditya_Blindspot}, the association cells  generated by $\Phi_{\rm m}$ and $\Phi_{\rm s}$ according to the association rules given by \eqref{eq::association::policy} and \eqref{eq::association::rule::backhaul}  are stationary partition of $\R^2$~\cite{Singh_association_cell}. Hence it is possible to characterize the mean area of a typical association cell (denoted by ${\cal C}_{{\rm a}_i}({\bf 0})$ and ${\cal C}_{\rm b}({\bf 0})$). 
\end{remark}
This helps us to formulate the following Proposition. 
\begin{prop}\label{prop::weightedVoronoi::access} The mean area of a typical  backhaul association cell is $\nbbE[|{\cal C}_{{\rm b}}({\bf 0})|] = \frac{1}{\lambda_{\rm m}}$
 and mean area of a typical access association cell is  $\nbbE[|{\cal C}_{{\rm a}_{i}}({\bf 0})|] = \frac{{\cal A}_{i}}{\lambda_{i}}$ ($i\in\{{\rm m},{\rm s}\}$). 
\end{prop}  
\begin{IEEEproof} 
Following Remark~\ref{rem::stationary},  \eqref{eq::association::rule::backhaul} is a stationary association strategy, for which  $\nbbE[|{\cal C}_{{\rm b}}({\bf 0})|] = \frac{1}{\lambda_{\rm m}}$~\cite{Singh_association_cell}. For the access association cells, a typical cell belongs to $\Phi_i$ with probability $\nbbP({\bf 0}\in\Phi_i) = {\cal A}_i$ (according to Definition~\ref{def::association::prob}). Hence, 
$
\nbbE[|{\cal C}_{{\rm a}_{i}}({\bf 0})|] = \frac{{\cal A}_i}{\lambda_i}$.
\end{IEEEproof}
We now explain the reason of calibrating $\mu$ according to Remark~\ref{rem::fitting::mu}. 
\begin{remark}\label{rem::simulation} 
According to Proposition~\ref{prop::weightedVoronoi::access},  we need  ${\cal A}_i$ to characterize the mean access association cell areas. 
While ${\cal A}_i$ can be evaluated analytically for simpler blockage models~\cite{SinghKulkarniSelfBackhaul}, its analytical characterization does not seem straightforward for the germ grain-model considered in Section~\ref{sec::sys::mod}. {
We empirically evaluate ${\cal A}_i$ for the typical user using the Matlab script provided in \cite{SahaIABCode}. Note that this empirical evaluation is way simpler than the ``true'' simulation of the load variables for which we need to drop multiple users and compute the link states of all possible links in the network.}
\end{remark}
We now compute the load distributions by assuming that the load PMF follows the same kernel as that of the load PMFs in sub-6 GHz networks~\cite{OffloadingSingh} with the mean loads for this case given by Proposition~\ref{prop::weightedVoronoi::access}. 
\begin{lemma} \label{lemm::access::load}
Under Proposition~\ref{prop::weightedVoronoi::access},  the PMFs of $\Phi_{\rm u}({\cal C}_{{\rm a}_{\rm m}}({\bf x}^*))$ and $\Phi_{\rm u}({\cal C}_{{\rm a}_{\rm s}}({\bf x}^*))$ are given as:
\begin{align}\label{eq::pmf::access::load}
 \nbbP(\Phi_{\rm u}({\cal C}_{{\rm a}_{i}}({\bf x}^*))= n ) \approx {\tt K}_t\bigg(n;\frac{\lambda_{i}}{{\cal A}_{i}},\lambda_{\rm u}\bigg),
  \nbbP(\Phi_{\rm u}({\cal C}_{{\rm a}_{\rm m}}(\tilde{\bf x}))= n ) \approx {\tt K}\bigg(n;\frac{\lambda_{i}}{{\cal A}_{\rm m}},\lambda_{\rm u}\bigg),\ i\in\{{\rm m},{\rm s}\},
\end{align}
where 
\begin{equation}\label{eq::Kt}
{\tt K}_t(n;\lambda,\lambda_{\rm u}) =\frac{3.5^{3.5}}{(n-1)!}\frac{\Gamma(n+3.5)}{\Gamma(3.5)}\left(\frac{\lambda_{\rm u}}{\lambda}\right)^{n-1}\left(3.5+\frac{\lambda_{\rm u}}{\lambda}\right)^{-n-3.5}, n\geq 1,
\end{equation}
 \begin{align}
 \nbbP(\Phi_{\rm s}({\cal C}_{\rm b}({\bf x}^*))=n) 
 \approx \nbbP(\Phi_{\rm s}({\cal C}_{\rm b}(\tilde{\bf x})=n) = {\tt K}(n;\lambda_{\rm m},\lambda_{\rm s}),\ n\geq 0,
\end{align}
and 
\begin{equation}\label{eq::K}
{\tt K}(n;\lambda,\lambda_{\rm u}) =\frac{3.5^{3.5}}{n!}\frac{\Gamma(n+3.5)}{\Gamma(3.5)}\left(\frac{\lambda_{\rm u}}{\lambda}\right)^{n}\left(3.5+\frac{\lambda_{\rm u}}{\lambda}\right)^{-n-3.5}, n\geq 0.
\end{equation}
\end{lemma}
Note that a random variable following the PMF ${\tt K}(\cdot;\lambda_1,\lambda_2)$ has mean $\lambda_2/\lambda_1$. 
\subsection{$\sinr$ Distributions}\label{subsec::sinr::distribution}
In this Section, we are going to evaluate the following CDFs: (i) {\em MBS coverage:} $\P(\sinr_{\rm a}({\bf 0})>\tau|{\bf x}\in\Phi_{\rm m})$, and (ii) {\em joint SBS and backhaul coverage:} $\P(\sinr_{\rm a}({\bf 0})>\tau_1,\sinr_{\rm b}({\bf x}^*)>\tau_2|{\bf x}\in\Phi_{\rm s})$. As noted earlier, the  germ-grain model for blockages  introduced in Section~\ref{sec::sys::mod} is not conductive for the $\sinr$ analysis~\cite{Bai_mmWave}. However, if $L_{\rm bl}$ is  not large enough,  we can characterize the $\sinr$ distributions by  {adopting the state-of-the-art}  {\em independent blocking model}~\cite{Aditya_Blindspot} which is stated as follows. 
\begin{assumption}\label{assumption::independent::blocking}
Each link state is assumed to be in LOS  independently   
 with  probability $p(r) = \exp(-r/\mu)$, where $r>0$ is the link distance and $\mu$ is the LOS range constant.
\end{assumption}
As evident in the sequel, the independent exponential blocking model closely approximates the germ-grain model of blocking in terms of the $\sinr$ distributions. 
 The connection between $\Phi_{\rm bl}$ and $\mu$ will be established in Remark~\ref{rem::fitting::mu}.  
 For preserving the simplicity of analysis, we now  make another reasonable assumption on the  $\sinr$.
\begin{assumption}\label{assumption::sinr::snr}(a) For the typical access link,  interference from the MBSs is neglected, i.e., 
\begin{equation}
{\tt SINR}_{\rm a}({\bf 0}) = \frac{P_i G_{i} G_{\rm u} \beta_{{\rm a}_{\rm i}} h_{0,{\bf x}^*}L_{{\rm a}_i}(\|{\bf x}^*\|)^{-1}}{\sum\limits_{{\bf x}\in\Phi_{{\rm s}}\setminus\{{{\bf x}}^*\}}{P_{\rm s}} h_{0,{\bf x}} \beta_{{\rm a}_{\rm s}}{\psi}_{{\rm a}_{\rm s}} L_{{\rm a}_{\rm s}}(\|{\bf x}\|)^{-1}+{\tt N}_0W},
\end{equation}
and (b) the tagged backhaul link is assumed to be noise-limited, i.e.,
\begin{equation}
\sinr_{\rm b} = \snr_{\rm b} = {\tt SINR}_{\rm b}({\bf x}^*) = \frac{P_{\rm m} G_{{\rm m}}G_{\rm s}\beta_{{\rm b}_{\rm m}} h_{{\bf x}^*,\tilde{\bf x}}L_{{\rm b}_{\rm m}}(\|{\bf x}^*-\tilde{\bf x}\|)^{-1}}
{{\tt N}_0W}.
\end{equation}
\end{assumption}
The intuition behind the above simplification is as follows. The SBSs, equipped with large antenna arrays, are able to beamform towards the direction of the ABS antenna to establish the backhaul link. On the other hand, the UEs, with lower    beamforming capabilities compared to the BSs, are likely to experience SBS interference due to the dense deployments alongside thermal noise.  
 As  will be clear in the  sequel, this reasonable assumption allows us to compute the joint distribution of $\sinr_{\rm a}({\bf 0})$ and $\sinr_{\rm b}({\bf x}^*)$, which is currently an open problem in the literature. 

As a first step towards the coverage and rate analyses, we define the pathloss point process on similar lines of \cite{direnzo2015stochastic,SinghKulkarniSelfBackhaul,mmWaveHetNetTurgut}. However note that due to the exponential blocking model considered in this paper, the properties of this process  are different than those of the LOS-ball model used in the prior arts (see \cite[Section III-C]{AndrewsMMWave} for details on the LOS-ball model). 
\begin{ndef}[Pathloss process]\label{def::pathloss::process} We define the sequence  $\{{\cal L}_{{k}_{i}}=L_{{k}_i}(\|{\bf x}\|):{\bf x}\in\Phi_{i}\}$ as a pathloss process  associated with $\Phi_{i}$ ($i\in \{{\rm m},{\rm s}\}$) where the reference point at the origin is the typical UE for $k = {\rm a}$ (corresponding to the typical access link) and typical SBS for $k={\rm b}$ (corresponding to the typical backhaul link). 
\end{ndef}
\begin{lemma} 
The pathloss process ${\cal L}_{{\rm k}_{i}}$ ($k\in\{{\rm a},{\rm b}\}$, $i\in\{{\rm m},{\rm s}\}$) 
is a PPP in  $\R^+$ with intensity measure: $\Lambda_{{k}_{i}}([0,l)) = $
\begin{multline}
2\pi\lambda_i\bigg[
\mu\bigg(
 \mu-e^{-\frac{l^{\frac{1}{\alpha_{{k}_{i,\l}}}}}{\mu}}
 \left(l^{\frac{1}{\alpha_{{k}_{i,\l}}}}+\mu\right)\bigg)
+\frac{l^{\frac{2}{\alpha_{{k}_{i,\n}}}}}{2}-\mu\bigg(\mu-e^{
-\frac{l^{\frac{1}{\alpha_{{k}_{i,\n}}}}}{\mu}}
\bigg(l^{\frac{1}{\alpha_{{k}_{i,\n}}}}+\mu\bigg)\bigg)\bigg],
\label{eq::Lambda}
\end{multline}
and density:
\begin{equation}\label{eq::Lambda_k_derivative}
\lambda_{{ k}_{i}}(l) = 2\pi\lambda_i\left(
\frac{1}{\alpha_{{k}_{i,\l}}}l^{\frac{2}{\alpha_{{k}_{i,\l}}}-1}e^{-\frac{l}{{\mu}}}+
\frac{1}{\alpha_{{k}_{i,\n}}}l^{\frac{2}{\alpha_{{k}_{i,\n}}}-1}\left(1-e^{-\frac{l}{{\mu}}}\right)\right),\ \text{for } l>0.
\end{equation}
\end{lemma}
\begin{IEEEproof}
Since the link state (i.e. LOS or NLOS) can be considered as independent mark on each BS in $\Phi_i$, the LOS and NLOS BSs with respect to the typical UE (typical SBS) are inhomogeneous PPPs with densities $\lambda_{i}e^{-\|{\bf x}\|/\mu}$ and $\lambda_i(1-e^{-\|{\bf x}\|/\mu})$, respectively~\cite{baccelli2009stochastic}. These PPPs under mapping $L_{{k}_{i}}(\|{\bf x}\|)$  are PPPs in $\R^+$. Superposition of these PPPs gives us ${\cal L}_{{k}_i}$, which is again a PPP with intensity measure:
\begin{multline*}
\Lambda_{{k}_{i}}([0,l))=2\pi\lambda_i \int\limits_{0}^{\infty}\nbbP(L_{{k}_{i}}(r)<l)r{\rm d}r=2\pi\lambda_i\int\limits_{0}^{\infty}\bigg(e^{-\frac{r}{\mu}}{\bf 1}(r^{\alpha_{{k}_{i,\l}}}<l)+\left(1-e^{-\frac{r}{\mu}}\right)\times\\{\bf 1}(r^{\alpha_{{k}_{i,\n}}}<l)\bigg)r{\rm d}r=
2\pi\lambda_i\int\limits_{0}^{l^{\frac{1}{\alpha_{k_{i,\l}}}}}
e^
{-\frac{r}{\mu}}r{\rm d}r
+2\pi\lambda_i\int\limits_0^{l^{\frac{1}{\alpha_{k_{i,\n}}}}}
\left(1-e^{-\frac{r}{\mu}}\right)r{\rm d}r.
\end{multline*}
The final expression in \eqref{eq::Lambda} follows from algebraic simplifications. Differentiating with respect to $l$ using Leibniz integral rule, we obtain the density function in \eqref{eq::Lambda_k_derivative}.  
\end{IEEEproof}
In the following Corollary, we provide the expressions of the intensity measures and densities of the PPPs formed by the LOS and NLOS BSs of $\Phi_{\rm s}$ with respect to the typical UE, to be later used for the derivation of the joint SBS and backhaul coverage. 
\begin{cor}\label{cor::pathloss::process::special::case}
The pathloss processes of the LOS and NLOS links from the BSs in $\Phi_{\rm s}$ to the typical UE are PPPs with intensity measures 
\begin{align}
&\Lambda_{{{\rm a}_{\rm s},\l}}([0,l)) = 
2\pi\lambda_{\rm s}\mu\left(\mu-e^{-\frac{l^{\frac{1}{\alpha_{{\rm a}_{{\rm s},\l}}}}}{\mu}}\left(l^{\frac{1}{\alpha_{{\rm a}_{{\rm s},\l}}}}+\mu\right)\right),\label{eq::Lambda::macro::los}\\
&\Lambda_{{{\rm a}_{\rm s},{\n}}}([0,l)) =2\pi\lambda_{\rm s}\left(\frac{l^{\frac{2}{\alpha_{{\rm a}_{{\rm s},\n}}}}}{2}-\mu\left(\mu-e^{
-\frac{l^{\frac{1}{\alpha_{{\rm a}_{{\rm s},\n}}}}}{\mu}}
\left(l^{\frac{1}{\alpha_{{\rm a}_{{\rm s},\n}}}}+\mu\right)\right)\right),
\label{eq::Lambda::macro::nlos}
\end{align} 
and density functions 
\begin{align}
&\lambda_{{\rm a}_{\rm s},\l}(l) = \frac{2\pi\lambda_{\rm s}
}{\alpha_{{\rm a}_{{\rm s},\l}}}l^{2/\alpha_{{\rm a}_{{\rm s},\l}}-1}e^{-\frac{l}{{\mu}}
},\ 
\lambda_{{\rm a}_{\rm s},\n}(l) = \frac{2\pi\lambda_{\rm s}}{\alpha_{{\rm a}_{{\rm s},\n}}}l^{2/\alpha_{{\rm a}_{{\rm s},\n}}-1}\left(1-e^{-\frac{l}{{\mu}}}\right),& l>0.
\end{align}
\end{cor}
Note that $\Lambda_{{\rm a}_{\rm s},\l}((0,l])+\Lambda_{{\rm a}_{\rm s},\n}((0,l])=\Lambda_{{\rm a}_{\rm s}}((0,l])$ and $\lambda_{{\rm a}_{\rm s},\l}(l)+\lambda_{{\rm a}_{\rm s},\n}(l)=\lambda_{{\rm a}_{\rm s}}(l)$.
Since the user association directly depends on the pathloss (see \eqref{eq::associaiton::cell::access}), we are now in a position to characterize the association probabilities to $\Phi_{\rm m}$ and $\Phi_{\rm s}$ for the typical access link defined in Definition~\ref{def::association::prob}. 
\begin{lemma} \label{lemm::association::prob}The association probability of the typical access link to a BS of $\Phi_{i}$ is expressed as
\begin{align}\label{eq::association::probability::access}
{\cal A}_{i} = \int\limits_{0}^{\infty}e^{-\sum\limits_{j\in{\{{\rm m},{\rm s}}\}}\Lambda_{{\rm a}_{j}}((0,{\Omega}_{j,i}l])}\lambda_{{\rm a}_{i}}(l){\rm d}l,
\end{align}
where $\Lambda_{{\rm a}_{j}}(\cdot)$ and $\lambda_{{\rm a}_{j}}(\cdot)$  are given by \eqref{eq::Lambda} and \eqref{eq::Lambda_k_derivative}, respectively, and $\Omega_{j,i} = \frac{P_jT_j\beta_{{\rm a}_j}G_j}{P_iT_i\beta_{{\rm a}_i}G_i}$.
\end{lemma} 
\begin{IEEEproof}
First we denote the location of candidate serving BS of $\Phi_{i}$ as 
\begin{equation*}
\bar{{\bf x}}_{i} = \arg\max\limits_{{\bf x}\in\Phi_{i}} P_iT_i\beta_{{\rm a}_i}G_{i}G_{\rm u}L_{{\rm a}_{i}}(\|{\bf x}\|)^{-1} = \arg\min\limits_{{\bf x}\in\Phi_i}L_{{\rm a}_i}({\bf x}).
\end{equation*}
The CDF of $\bar{L}_{{\rm a}_i}:=L_{{\rm a}_{i}}(\bar{\bf x}_{i})$ can be obtained from the CDF of the contact distance of ${\cal L}_{{\rm a}_{i}}$ as
$
\nbbP(\bar{L}_{{\rm a}_i} \leq l)= 1-e^{-\Lambda_{{\rm a}_{i}}((0,l])}.
$ 
Differentiating with respect to $l$, we obtain the PDF of $\bar{L}_{{\rm a}_i}$ as:
$f_{\bar{L}_{{\rm a}_i}}(l) =  e^{-\Lambda_{{\rm a}_{i}}((0,l])}\lambda_{{\rm a}_{i}}(l),
 l >0.$
 Now, ${\cal A}_{i} = \nbbP({\bf x}^*\in\Phi_{i}) =\nbbP\big(P_iT_i\beta_{{\rm a}_i}G_{i}\bar{L}_{{\rm a}_{i}}^{-1}\geq P_jT_j\beta_{{\rm a}_j}G_{j}\bar{L}_{{\rm a}_{j}}^{-1}\big)
{=} \nbbP\big(\bar{L}_{{\rm a}_{j}}\geq \Omega_{j,i}\bar{L}_{{\rm a}_{i}}\big) =
\nbbE\big[e^{-\Lambda_{{\rm a}_{j}}((0,{\Omega}_{j,i}\bar{L}_{{\rm a}_{i}}])}\big]
=\int\limits_{0}^{\infty}e^{-\Lambda_{{\rm a}_{j}}((0,{\Omega}_{j,i}l])}f_{\bar{L}_{{\rm a}_{i}}}(l){\rm d}l$.  The final expression is obtained by substituting $f_{\bar{L}_{{\rm a}_i}}(l)$. 
\end{IEEEproof}
We remind that the results derived in this Section are functions of $\mu$,  which appears in the expression for the blocking probability given in Assumption~\ref{assumption::independent::blocking}. While characterizing $\mu$ for the germ-grain blockage model of Section~\ref{sec::sys::mod} is known to be hard, we propose one reasonable  way of choosing $\mu$ given a particular blockage configuration ($\lambda_{\rm bl}$, $L_{\rm bl}$) in the following remark. 
\begin{remark}\label{rem::fitting::mu}
We choose $\mu$ such that ${\cal A}_i$ in  \eqref{eq::association::probability::access} evaluated as a function of $\mu$ is equal to the empirical value of ${\cal A}_i$ computed as Definition~\ref{def::association::prob}. Since we have a two-tier network, it is sufficient to match only one quantity, say, ${\cal A}_{\rm m}$ (since ${\cal A}_{\rm s}=1-{\cal A}_{\rm m}$) for the calibration of $\mu$. A simple Matlab script to empirically obtain the value of $\mu$ is provided by the authors at~\cite{SahaIABCode}.
\end{remark}
While one can of course use other ways to calibrate $\mu$ with the blockage parameters~\cite{AndrewsMMWave}, the reason of this  particular way of calibration will be clarified in the next Section.
We now derive the distribution of pathloss of the serving link, i.e., the link between the typical UE and its serving BS. 
\begin{lemma}\label{lemm::serving::dist::distribution}
Conditioned on ${\bf x}^*\in\Phi_i$, the PDF of $L^*_{\rm a} :=L_{{\rm a}_i}(\|{\bf x}^*\|)$ is given by
\begin{align}\label{eq::serving::dist::distribution} 
f_{L^*_{\rm a}}(l|{\bf x}^*\in\Phi_i)=\frac{1}{{\cal A}_{i}}e^{-\sum\limits_{j\in{\{{\rm m},{\rm s}}\}}\Lambda_{{\rm a}_{j}}((0,{\Omega}_{j,i}l])}\lambda_{{\rm a}_{i}}(l),\qquad l>0,
\end{align}
where $\Lambda_{{\rm a}_{j}}(\cdot)$, $\lambda_{{\rm a}_{j}}(\cdot)$, and ${\cal A}_i$ are given by  \eqref{eq::Lambda}, \eqref{eq::Lambda_k_derivative}, and \eqref{eq::association::probability::access}, respectively.
\end{lemma}
\begin{IEEEproof}The conditional CCDF of $L^*_{\rm a}$ given ${\bf x}^*\in\Phi_{i}$ is $\bar{F}_{L^*_{\rm a}}(l|{\bf x}^*\in\Phi_i) = $
\begin{align*}
&\nbbP\left(\bar{L}_{{\rm a}_i}>l|{\bf x}^*\in\Phi_{i}\right)=\frac{\nbbP\left(\bar{L}_{{\rm a}_i}>l,{\bf x}^*\in\Phi_{i}\right)}{\nbbP\left({\bf x}^*\in\Phi_{i}\right)}=\frac{1}{{\cal A}_i}\int\limits_{l}^{\infty}e^{-\sum\limits_{j\in{\{{\rm m},{\rm s}}\}}\Lambda_{{\rm a}_{j}}((0,{\Omega}_{j,i}l])}\lambda_{{\rm a}_{i}}(l){\rm d}l.
\end{align*}
The desired PDF can be obtained by differentiating with respect to $l$. 
\end{IEEEproof}
In Lemmas~\ref{lemm::association::prob}-\ref{lemm::serving::dist::distribution}, we derived the association probability and pathloss PDFs of the serving link for the two-tier HetNet. One can further interpret this network as a three-tier HetNet by splitting $\Phi_{\rm s}$ into  $\Phi_{{\rm s},\l}$ and $\Phi_{{\rm s},\n}$ which are the sets of   SBSs at LOS and NLOS of the typical UE, respectively. In the following Corollary, we provide the association probabilities to $\Phi_{{\rm s},\l}$ and $\Phi_{{\rm s},\n}$ and the corresponding PDFs of pathloss of the serving link. 
\begin{cor}\label{cor::pathloss::special::case::association::servingPDF}
 The SBS association event can split into two events based on the state of the link between the typical UE and serving SBS:
\begin{align*}
{\cal A}_{\rm s} = \underbrace{\nbbP({\bf x}^*\in\Phi_{\rm s},s({\bf x}^*,{\bf 0})=\l)}_{{\cal A}_{{\rm s}_\l}}+ \underbrace{\nbbP({\bf x}^*\in\Phi_{\rm s},s({\bf x}^*,{\bf 0})=\n)}_{{\cal A}_{{\rm s}_\n}},
\end{align*}
where the association probabilities to LOS SBS and NLOS SBS are given by
\begin{align*}
{\cal A}_{{\rm s}_{t}} = \int\limits_{0}^{\infty}e^{-\sum\limits_{j\in{\{{\rm m},{\rm s}}\}}\Lambda_{{\rm a}_{j}}((0,{\Omega}_{j,{\rm s}}l])}\lambda_{{\rm a}_{{\rm s},t}}(l){\rm d}l, \qquad t\in\{\l,\n\},
\end{align*}
and the corresponding pathlosses of the serving links are denoted as $L_{\rm a}^*|s({\bf x}^*,{\bf 0})=\l$ and $L_{\rm a}^*|s({\bf x}^*,{\bf 0})=\n$, respectively whose PDFs are given as
\begin{align}
&f_{L^*_{\rm a}}(l|{\bf x}^*\in\Phi_{\rm s},s({\bf x}^*,{\bf 0})=t) =\frac{1}{{\cal A}_{{\rm s}_{t}}}e^{-\sum\limits_{j\in{\{{\rm m},{\rm s}}\}}\Lambda_{{\rm a}_{j}}((0,{\Omega}_{j,{\rm s}}l])}\lambda_{{\rm a}_{{\rm s},t}}(l),\qquad l>0,t\in\{\l,\n\},\label{eq::pathloss::pdf::sbs}
\end{align}
where $\Lambda_{{\rm a}_{{\rm s}},\l},\Lambda_{{\rm a}_{{\rm s}},\n},\lambda_{{\rm a}_{{\rm s}},\l},$ and $\lambda_{{\rm a}_{{\rm s}},\n}$ are given by Corollary~\ref{cor::pathloss::process::special::case}. 
\end{cor}
\begin{IEEEproof}
The SBS PPP can be treated as a superposition of LOS and NLOS SBS PPPs. Considering these two PPPs instead of $\Phi_{\rm s}$, the proof follows on similar lines of Lemmas~\ref{lemm::association::prob} and \ref{lemm::serving::dist::distribution}  for a three-tier HetNet.
\end{IEEEproof}
We now characterize the pathloss process of MBSs for the {\em tagged} backhaul link which  is not the same as that of the {\em typical} backhaul link  since it is conditioned on the pathloss of the typical access link ($L_{\rm a}^*$)  and the fact that ${\bf x}^*\in\Phi_{\rm s}$. Note that this pathloss characterization is required to derive the joint SBS and backhaul coverage.
\begin{lemma}\label{lemm::joint::pathloss::distribution}
The pathloss process formed by the MBSs perceived by the tagged SBS at ${\bf x}^*$   conditioned on the pathloss and state of the typical access link i.e. $L_{\rm a}^*$, $s({\bf x}^*,{\bf 0})$, the location of the serving SBS at ${\bf x}^* \equiv ({L_{\rm a}^*}^{1/\alpha_{{\rm a},s({\bf x}^*,{\bf 0})}},\theta^*) $, and the association to SBS (${\bf x}^*\in\Phi_{\rm s}$), denoted as  
 ${\cal L}_{{\rm b}_{\rm m}}| s({\bf x}^*,{\bf 0}),{\bf x}^*\in\Phi_{\rm s},{\bf x}^* \equiv ({L_{\rm a}^*}^{1/\alpha_{{\rm a},s({\bf x}^*,{\bf 0})}},\theta^*)$ are PPPs in $\R^+$ with intensity measure
\begin{align}
&\tilde{\Lambda}_{{\rm b}_{t}}((0,l];L_{\rm a}^*,\theta^*)=
\int_0^{2\pi}\left(
\int_0^{l^{1/\alpha_{{\rm b}_{{\rm m},\l}}}}\tilde{\lambda}_{\rm m}(r,\theta;{L_{\rm a}^*}^{1/\alpha_{{\rm a}_{{\rm s},t}}},\theta^*)e^{-r/\mu}r\:{\rm d}r \right.\notag\\
&\left.+\int_0^{l^{1/\alpha_{{\rm b}_{{\rm m},\n}}}}\tilde{\lambda}_{\rm m}(r,\theta;{L_{\rm a}^*}^{1/\alpha_{{{\rm a}_{{\rm s},t}}}},\theta^*)(1-e^{-r/\mu})r\:{\rm d}r\right)\:{\rm d}\theta,\quad\text{ for }t= s({\bf x}^*,{\bf 0}) \in \{\l,\n\},\label{eq::Lambda_b_l}
\end{align}
and  intensity function 
\begin{align}
&\tilde{\lambda}_{{\rm b}_{t}}(l;L_{\rm a}^*,\theta^*) = \int_0^{2\pi}\tilde{\lambda}_{\rm m}(l^{1/\alpha_{{\rm b}_{{\rm m},\l}}},\theta;{L_{\rm a}^*}^{1/\alpha_{{\rm a}_{{\rm s},t}}},\theta^*)
\frac{1}{\alpha_{{\rm b}_{{\rm m},\l}}}l^{\frac{2}{\alpha_{{\rm b}_{{\rm m},\l}}}-1}e^{-\frac{l}{\mu}}+
\tilde{\lambda}_{\rm m}(l^{1/\alpha_{{\rm b}_{{\rm m},\n}}},\theta;{L_{\rm a}^*}^{1/\alpha_{{\rm a}_{{\rm s},t}}},\theta^*)\times\notag\\&
\frac{1}{\alpha_{{\rm b}_{{\rm m},\n}}}l^{\frac{2}{\alpha_{{\rm b}_{{\rm m},\n}}}-1}\left(1-e^{-\frac{l}{\mu}}\right)\:{\rm d}{\theta},\ \text{for }t=s({\bf x}^*, {\bf 0})\in\{\l,\n\},\label{eq::lambda_b_l}
\end{align}
where $
\tilde{\lambda}_{\rm m}(r, \theta;x,\theta^*) = 
\tilde{\lambda}_{\rm m}'(r^2+x^{2}-2rx\cos(\theta-\theta^*))^{\frac{1}{2}},
$
with 
$
\tilde{\lambda}_{\rm m}'(r)= 
\lambda_{\rm m}e^{-r/\mu}
{\bf 1}
\left(
 r>\left(\Omega_{{\rm s},{\rm m}}{L_{\rm a}^*}\right)^{\frac{1}{\alpha_{{\rm a}_{{\rm m},\l}}}}\right)+ \lambda_{\rm m}
 (1-e^{-r/\mu}){\bf 1}\left(r>\left(\Omega_{{\rm s},{\rm m}}{L_{\rm a}^*}\right)^{\frac{1}{\alpha_{{\rm a}_{{\rm m},\n}}}}\right).
$
\end{lemma}
\begin{IEEEproof}The point process $\Phi_{\rm m}| \{L_{\rm a}^*, {\bf x}^*\in\Phi_{\rm s}, s({\bf x}^*,{\bf 0})={\l}\}$ is a PPP in $\mathbb{R}^2$ with density:
\begin{equation*}
\tilde{\lambda'}_{\rm m}(r)= \lambda_{\rm m} e^{-r/\mu}{\bf 1}\left(r>\Omega_{{\rm s},{\rm m}}^{\frac{1}{\alpha_{{\rm a}_{{\rm m},\l}}}}{L_{\rm a}^*}^{\frac{1}{\alpha_{{\rm a}_{{\rm m},\l}}}}\right)+\lambda_{\rm m} (1-e^{-r/\mu}){\bf 1}\left(r>\Omega_{{\rm s},{\rm m}}^{\frac{1}{\alpha_{{\rm a}_{{\rm m},\n}}}}{L_{\rm a}^*}^{\frac{1}{\alpha_{{\rm a}_{{\rm m},\n}}}}\right).
\end{equation*}
 When this point process is seen from the tagged SBS at ${\bf x}^* = ({L_{\rm a}^*}^{\frac{1}{\alpha_{{\rm s},\l}}},\theta^*)$, the density becomes $\tilde{\lambda}_{\rm m}(r,\theta;{\bf x}^*) = \tilde{\lambda'}_{\rm m}((r^2+{L_{\rm a}^*}^{\frac{2}{\alpha_{{\rm s},\l}}}-2r{L_{\rm a}^*}^{\frac{1}{\alpha_{{\rm s},\l}}}\cos(\theta-\theta^*))^{\frac{1}{2}})$. Now the pathloss process on $\R^+$ for this conditional version of $\Phi_{\rm m}$ perceived by the tagged SBS will be a PPP with intensity function: $\tilde{\Lambda}_{{\rm b}_{\l}}((0,l]|{\bf x}^*) = \int_0^{\infty}\int_0^{2\pi} \tilde{\lambda}_{\rm m}(r,\theta;{\bf x}^*)\mathbb{P}(L_{{\rm b}_{\rm m}}(r)<l){\rm d}\theta\: r\:{\rm d}r = $
   \begin{align*}
&\int_0^{\infty}\int_0^{2\pi}\tilde{\lambda}_{\rm m}(r,\theta;{\bf x}^*)\bigg(e^{-r/\mu}{\bf 1}(r^{\alpha_{{\rm b}_{{\rm m},\l}}}<l)+(1-e^{-r/\mu}){\bf 1}(r^{\alpha_{{\rm b}_{{\rm m},\n}}}<l)\bigg)\:{\rm d}\theta\:r\:{\rm d}r.
\end{align*}
Differentiating with respect to $l$, we obtain the intensity function. Similar steps can be followed when $s({\bf x}^*,{\bf 0})= \n$.  
\end{IEEEproof}
We now obtain the $\sinr$ CCDFs required for the rate analysis as follows. 
\begin{lemma}\label{lemm::joint::coverage}
The MBS and the joint SBS and backhaul coverages  under Assumption~\ref{assumption::sinr::snr} is given by
\begin{align}
&\nbbP(\sinr_{\rm a}({\rm 0})>\tau|{\bf x}^*\in\Phi_{\rm m})  =  \frac{1}{{\cal A}_{\rm m}}
 \int\limits_{0}^{\infty}
\exp\left(-\sum\limits_{{\cal G}\in{\cal M}_{{\rm a}_{\rm s}}}\int\limits_{\Omega_{{\rm s},{\rm m}}l}^{\infty}\left(1-\frac{1}{1+\frac{\tau P_{\rm s}\beta_{{\rm a}_{\rm s}}{\cal G} l}{P_{\rm m}\beta_{{\rm a}_{\rm m}}G_{\rm m}G_{\rm u}z}}\right)p_{{\cal G}}\lambda_{{\rm a}_{\rm s}}(z){\rm d}z\right.\notag\\&\left.\qquad\qquad\qquad-\frac{\tau{\tt N}_0Wl}{P_{\rm m}\beta_{{\rm a}_{\rm m}}G_{\rm m}G_{\rm u}}-\sum\limits_{j\in{\{{\rm m},{\rm s}}\}}\Lambda_{{\rm a}_{j}}((0,{\Omega}_{j,{\rm m}}l])\right)\lambda_{{\rm a}_{\rm m}}(l)
{\rm d}l,\label{eq::sinr::macro::coverage}\\
&\nbbP({\sinr}_{\rm a}({\bf 0})>\tau_1,\snr_{\rm b}({\bf x}^*)>\tau_2|{\bf x}^*\in\Phi_{\rm s})=\frac{1}{{\cal A}_{\rm s}}
\sum\limits_{t\in\{l,n\}}
 \int\limits_{0}^{\infty}\int\limits_{0}^\infty
\exp\left(-\sum\limits_{{\cal G}\in{\cal M}_{{\rm a}_{\rm s}}}\int\limits_{l_1}^{\infty}\left(1-\frac{1}{1+\frac{\tau_1 {\cal G} l_1}{G_{\rm s}G_{\rm u}z}}\right)\times\right.\notag\\&\left.p_{\cal G}\lambda_{{\rm a}_{\rm s}}(z){\rm d}z
-\frac{\tau_1{\tt N}_0Wl_1}{P_{\rm s}\beta_{{\rm a}_{\rm s}}G_{\rm s}G_{\rm u}}-\frac{\tau_2{\tt N}_0Wl_2}{P_{\rm m}\beta_{{\rm b}_{\rm m}}G_{\rm m}G_{\rm s}}-\tilde{\Lambda}_{{\rm b}_t}((0,l_2];l_1,0)-\sum\limits_{j\in{\{{\rm m},{\rm s}}\}}\Lambda_{{\rm a}_{j}}((0,{\Omega}_{j,{\rm s}}l_1])\right)\notag\\&\times\tilde{\lambda}_{{\rm b}_{{\rm s},t}}(l_2;l_1,0)\lambda_{{\rm a}_{{\rm s},t}}(l_1)\:{\rm d}l_2 \:{\rm d}l_1.\label{eq::joint::coverage::SBS}
\end{align}
\end{lemma}
\begin{IEEEproof}
See Appendix~\ref{app::lemm::joint::coverage}.
\end{IEEEproof}
Note that the summation appearing in the expression of joint SBS and backhaul coverage in \eqref{eq::joint::coverage::SBS} is over the link states of the access link between the typical user and the serving SBS. 
\begin{remark}
It is worth mentioning that one of the key contributions of this paper is the characterization of the joint SBS and backhaul coverage in mm-wave HetNets. This is enabled by the exponential path-loss assumption and  can facilitate the   analysis of joint coverage in other similar settings such as~\cite{krishnan2016spatio}, some of which may yield much simpler forms for the expression.
\end{remark}
However, for our case,  since \eqref{eq::joint::coverage::SBS} contains \eqref{eq::Lambda_b_l} and \eqref{eq::lambda_b_l} which have integrals over discontinuous functions that are prone to numerical errors, we simplify the expression of joint SBS and backhaul coverage with the following Assumption. As will be evident in the sequel, this facilitates further analysis without compromising the accuracy of the results and design insights.
\begin{assumption}\label{assumption::indenpendent-access-backhaul-snr} The joint SBS and backhaul coverage is approximated as the product of the coverages of a typical access and typical backhaul link:
\begin{equation}\label{eq::assumption::indenpendent-access-backhaul-snr}
\nbbP({\sinr}_{\rm a}({\bf 0})>\tau_1,\snr_{\rm b}({\bf x}^*)>\tau_2|{\bf x}^*\in\Phi_{\rm s})=
\nbbP({\sinr}_{\rm a}({\bf 0})>\tau_1|{\bf x}^*\in\Phi_{\rm s})\nbbP(\snr_{\rm b}({\bf 0})>\tau_2), 
\end{equation}
where $\snr_{\rm b}({\bf 0})$ is the $\snr$ of a typical backhaul link. 
\end{assumption}    
The main reason for the expression of \eqref{eq::joint::coverage::SBS} to be complex 
 is  the correlation of  $\sinr_{\rm a}({\bf x}^*)$  and $\snr_{\rm b}(\tilde{\bf x})$.  Since we ignore this correlation in the above assumption,  we obtain a simpler expression for the joint SBS and backhaul coverage in the following Corollary. 
\begin{cor}
\label{cor::joint::sbs::backhaul::coverage}
Under Assumption~\ref{assumption::indenpendent-access-backhaul-snr}, the joint SBS and backhaul coverage is given by:
\begin{multline}
\nbbP({\sinr}_{\rm a}({\bf 0})>\tau_1|{\bf x}^*\in\Phi_{\rm s})\nbbP(\snr_{\rm b}({\bf 0})>\tau_2) = \frac{1}{{\cal A}_{\rm s}}
 \int\limits_{0}^{\infty}
\exp\bigg(-\sum\limits_{{\cal G}\in{\cal M}_{{\rm a}_{\rm s}}}\int\limits_{l_1}^{\infty}\left(1-\frac{1}{1+\frac{\tau_1 {\cal G} l_1}{G_{\rm s}z}}\right)\times\\p_{\cal G}\lambda_{{\rm a}_{\rm s}}(z){\rm d}z
-\frac{\tau_1{\tt N}_0Wl_1}{P_{\rm m}\beta_{{\rm a}_{\rm m}}G_{\rm m}}-\sum\limits_{j\in{\{{\rm m},{\rm s}}\}}\Lambda_{{\rm a}_{j}}((0,{\Omega}_{j,{\rm s}}l_1])\bigg)\lambda_{{\rm a}_{{\rm s}}}(l_1) \:{\rm d}l_1\\
\times\int\limits_{0}^{\infty}\exp\left(-\frac{\tau_2{\tt N}_0Wl_2}{P_{\rm m}\beta_{{\rm b}_{\rm m}}G_{\rm m}G_{\rm s}}-{\Lambda}_{{{\rm b}_{\rm m}}}((0,l_2])\right){\lambda}_{{\rm b}_{{\rm m}}}(l_2)\:{\rm d}l_2 \label{eq::joint::coverage::approx}
\end{multline}
\end{cor}
\begin{IEEEproof} The two probability terms appearing in the product can be handled separately. The first term can be simplified by following the same steps used to derive the MBS coverage in Appendix~\ref{app::lemm::joint::coverage}. For the second term, $\nbbP(\snr_{\rm b}({\bf 0})>\tau_2) =$
\begin{align*}
&\nbbP\bigg( \frac{P_{\rm m}\beta_{{\rm a}_{\rm m}}G_{\rm m}G_{\rm s} h_{{\bf 0},\tilde{\bf x}}L_{{\rm b}_{\rm m}}(\|\tilde{\bf x}\|)^{-1}}{{\tt N}_0W}>\tau_2 \bigg)
=\nbbP\bigg(h_{{\bf 0},\tilde{\bf x}}> \frac{\tau_2 \tilde{L}_{{\rm b}_{\rm m}}{\tt N}_0W}{P_{\rm m}\beta_{{\rm a}_{\rm m}}G_{\rm m}G_{\rm s}} \bigg)=\nbbE\bigg[e^{-\frac{\tau_2 \tilde{L}_{{\rm b}_{\rm m}}{\tt N}_0W}{P_{\rm m}\beta_{{\rm a}_{\rm m}}G_{\rm m}G_{\rm s}} }\bigg],
\end{align*}
where $\tilde{{\bf x}}$ denotes the location of the MBS serving the typical SBS and $\tilde{L}_{{\rm b}_{\rm m}} = \min({\cal L}_{{\rm b}_{\rm m}})$, where the PPP ${\cal L}_{{\rm b}_{\rm m}}$ is defined in Definition~\ref{def::pathloss::process}.  Thus, the PDF of $\tilde{L}_{{\rm b}_{\rm m}}$ is given by
$
f_{\tilde{L}_{{\rm b}_{\rm m}}}(l) =  \lambda_{{\rm b}_{\rm m}}(l)e^{-\Lambda_{{\rm b}_{\rm m}}((0,l])}, \ l>0$. 
The final expression is obtained by deconditioning with respect to  $\tilde{L}_{{\rm b}_{\rm m}}$. Hence, $
\nbbP(\snr_{\rm b}({\bf 0})>\tau_2) = \int\limits_{0}^{\infty}  e^{-\frac{\tau_2 l_2 {\tt N}_0W}{P_{\rm m}\beta_{{\rm a}_{\rm m}}G_{\rm m}G_{\rm s}} }   \lambda_{{\rm b}_{\rm m}}(l_2)e^{-\Lambda_{{\rm b}_{\rm m}}((0,l_2])}{\rm d}l_2.
$
\end{IEEEproof}

\subsection{Rate Coverage Probability}
We are now in position to evaluate the rate coverage. 
\begin{theorem}\label{theorem::rate::coverage}
Rate coverage for a typical UE in the two-tier HetNet with IAB introduced in Section~\ref{sec::sys::mod} for a target rate-threshold $\rho$ is expressed as follows.
\begin{multline}\label{eq::rate::coverage::IRA}
   \pr^{\rm IRA}(\rho)  = {\cal A}_{\rm m}\sum\limits_{n=1}^{\infty}\nbbP\bigg(
 \sinr_{\rm a}({\bf 0})>
 2^
 {
  \frac{\rho}{W}
     \left(
      n+\frac{{\cal A}_{\rm s}{\lambda}_{\rm u}}{\lambda_{\rm m}}
     \right)
 }-1{\bigg|{\bf x}^*\in\Phi_{\rm m}}\bigg){\tt K}_t\left(n;\frac{\lambda_{\rm m}}{{\cal A}_{\rm m}},\lambda_{\rm u}\right) \\ + {\cal A}_{\rm s}\sum\limits_{n=1}^{\infty}\nbbP\bigg({\sinr}_{\rm a}({\bf 0})>
2^{\frac{\rho n}{W}\left(1+\frac{\lambda_{\rm m}n}{\lambda_{\rm u}}\right)}-1{\bigg|{\bf x}^*\in\Phi_{\rm s}}\bigg)\nbbP\bigg(\snr_{\rm b}({\bf 0})>2^{\frac{\rho\left(n+\frac{\lambda_{\rm u}}{\lambda_{\rm m}}\right)}{W}}-1\bigg) {\tt K}_t\left(n;\frac{\lambda_{\rm s}}{{\cal A}_{\rm s}},\lambda_{\rm u}\right).
\end{multline}  
\begin{multline}\label{eq::rate::coverage::ORA}
   \pr^{\rm ORA}(\rho)  ={\cal A}_{\rm m}\sum\limits_{n=1}^{\infty}\nbbP\bigg({\sinr}_{\rm a}({\bf 0})>2^{\frac{\rho n}{\eta_{\rm a} W }}-1{\bigg|{\bf x}^*\in\Phi_{\rm m}}\bigg){\tt K}_t\left(n;\frac{\lambda_{\rm m}}{{\cal A}_{\rm m}},{\lambda}_{\rm u}\right)\\+ {\cal A}_{\rm s}\sum\limits_{n=1}^{\infty}
   \nbbP({\sinr}_{\rm a}({\bf 0})>2^{\frac{\rho n}{W\eta_{\rm a}}}-1{|{\bf x}^*\in\Phi_{\rm s}})\nbbP(\snr_{\rm b}({\bf 0})>2^{\frac{\rho\left(n+\frac{{\cal A}_{\rm s}\lambda_{\rm u}}{\lambda_{\rm m}}\right)}{W(1-\eta_{\rm a})}}-1){\tt K}_t\left(n;\frac{\lambda_{\rm s}}{{\cal A}_{\rm s}},\lambda_{\rm u}\right), 
\end{multline}  
where the MBS coverage, {joint SBS and backhaul coverage}, and  ${\tt K}_t(\cdot)$ are given by \eqref{eq::sinr::macro::coverage},  \eqref{eq::joint::coverage::approx}, and  \eqref{eq::Kt}, respectively. 
\end{theorem}
\begin{IEEEproof}
For IRA, following \eqref{eq::Rate}, the CCDF of ${\tt Rate}^{{\rm IRA}}$ is given by
\begin{multline}
\nbbP({\tt Rate}_{\rm IRA}{>\rho}) = {\cal A}_{\rm m}\nbbP\bigg(
 \frac{W}{\Phi_{\rm u}({\cal C}_{{\rm a}_{\rm m}}({\bf x}^*))+\sum\limits_{\substack{{\bf x}\in\Phi_{\rm s}\cap {\cal C}_{\rm b}({\bf x}^*)}}\Phi_{\rm u}({\cal C}_{{\rm a}_{\rm s}}({\bf x}))}\log\left(1+\sinr_{\rm a}({\bf 0})\right)>\rho\bigg) \\+ {\cal A}_{\rm s}
\nbbP \bigg( \frac{W}{\Phi_{\rm u}({\cal C}_{{\rm a}_{\rm s}}({\bf x}^*))}\min\left(\omega\log\left(1+\sinr_{\rm b}({\bf x}^*)\right) ,(1-\omega)\log\left(1+\sinr_{\rm a}({\bf 0})\right)\right)>\rho\bigg).\label{eq::proof::rate::cov::iterm::p1}
\end{multline}
The first term of the summation, i.e., the conditional rate coverage when the typical UE connects to an MBS can be  simplified as ${\cal A}_{\rm m}\nbbP\bigg(
 \sinr_{\rm a}({\bf 0})>2^{\frac{\rho}{W}\left({\Phi_{\rm u}({\cal C}_{{\rm a}_{\rm m}}({\bf x}^*))+\sum\limits_{{{\bf x}\in\Phi_{{\rm a}_{\rm s}}\cap {\cal C}_{\rm b}({\bf x}^*)}}\Phi_{\rm u}({\cal C}_{{\rm a}_{\rm s}}({\bf x}))}\right)}-1\bigg).$
The PMF of $\Phi_{\rm u}({\cal C}_{\rm m}({\bf x}^*))$ is given by Lemma~\ref{lemm::access::load}. The second term can be approximated as  the average number of UEs per SBS (i.e. ${\cal A}_{\rm s}\lambda_{\rm u}/\lambda_{\rm s}$) times the number of SBSs falling in ${\cal C}_{\rm b}({\bf x}^*)$: 
$\sum_{{{\bf x}\in\Phi_{\rm s}\cap {\cal C}_{\rm b}({\bf x}^*)}}\Phi_{\rm u}({\cal C}_{{\rm a}_{\rm s}}({\bf x}))\approx  \frac{{\cal A}_{\rm s}{\lambda}_{\rm u}}{\lambda_{\rm s}}\nbbE[\Phi_{\rm s}({\cal C}_{\rm  b}({\bf x}^*))]= \frac{{\cal A}_{\rm s}{\lambda}_{\rm u}}{\lambda_{\rm s}}\times \frac{\lambda_{\rm s}}{\lambda_{\rm m}}= \frac{{\cal A}_{\rm s}{\lambda}_{\rm u}}{\lambda_{\rm m}}.
$
The PMF of $\Phi_{\rm s}({\cal C}_{\rm  b}({\bf x}^*))$ is given by Lemma~\ref{lemm::access::load}. We now focus on the second term  in \eqref{eq::proof::rate::cov::iterm::p1} which can be simplified  as:
\begin{multline*}
 {\cal A}_{\rm s}
\nbbP\bigg(\sinr_{\rm a}({\bf 0})>2^{\frac{\rho}{W}{\Phi_{\rm u}({\cal C}_{{\rm a}_{\rm s}}({\bf x}^*))}\left[1+\frac{\Phi_{\rm u}({\cal C}_{{\rm a}_{\rm s}}({\bf x}^*))}{\Phi_{\rm u}({\cal C}_{{\rm a}_{\rm m}}({\tilde{{\bf x}}}))+\sum_{{\bf x}\in \Phi_{\rm s}\cap {\cal C}_{\rm b}(\tilde{\bf x})\setminus\{{\bf x}^*\}}\Phi_{\rm u}({\cal C}_{{\rm a}_{\rm s}}({\bf x}))}\right]}-1\bigg)\times \\
\nbbP\bigg(\snr_{\rm b}({\bf 0})>2^{\frac{\rho}{W}\left(\Phi_{\rm u}({\cal C}_{{\rm a}_{\rm m}}({\tilde{{\bf x}}}))+\Phi_{\rm u}({\cal C}_{{\rm a}_{\rm s}}({{\bf x}^*}))+\sum_{{\bf x}\in \Phi_{\rm s}\cap {\cal C}_{\rm b}(\tilde{\bf x})\setminus\{{\bf x}^*\}}\Phi_{\rm u}({\cal C}_{{\rm a}_{\rm s}}({\bf x}))\right)}-1\bigg).
\end{multline*}
 To obtain the final expression, the following approximations on the load variables is applied:
$
\sum_{{\bf x}\in \Phi_{\rm s}\cap {\cal C}_{\rm b}(\tilde{\bf x})\setminus\{{\bf x}^*\}}\Phi_{\rm u}({\cal C}_{{\rm a}_{\rm s}}({\bf x})) \approx \frac{{\cal A}_{\rm s}\lambda_{\rm u}}{\lambda_{\rm s}}\nbbE[\Phi_{\rm s}({\cal C}_{\rm b}({\tilde{\bf x}}))]= \frac{{\cal A}_{\rm s}\lambda_{\rm u}}{\lambda_{\rm s}}\times\frac{\lambda_{\rm s}}{\lambda_{\rm m}}= \frac{{\cal A}_{\rm s}\lambda_{\rm u}}{\lambda_{\rm m}}, $ and  $\Phi_{\rm u}({\cal C}_{{\rm a}_{\rm m}}(\tilde{\bf x}))\approx \nbbE[\Phi_{\rm u}({\cal C}_{{\rm a}_{\rm m}}(\tilde{\bf x}))]= \frac{{\cal A}_{\rm m}\lambda_{\rm u}}{\lambda_{\rm m}}$.

For ORA, following \eqref{eq::Rate::ORA}, $\pr^{\rm ORA} = $
\begin{multline*}
{\cal A}_{\rm m}
\nbbP\left(
\sinr_{\rm a}({\bf 0})
>
2^{\frac{\rho\Phi_{\rm u}({\cal C}_{{\rm a}_{\rm m}}({\bf x}^*))}{\eta_{\rm a}W}}-1\right)
+{\cal A}_{\rm s}\nbbP\left(\sinr_{\rm a}({\bf 0})>2^{\frac{\rho\Phi_{\rm u}({\cal C}_{{\rm a}_{\rm s}}({\bf x}^*))}{\eta_{\rm a}W}}-1\right)\\
\times\nbbP\left(\snr_{\rm b}({\bf 0}) >
2^{\frac{
\sum_{{{\bf x}\in {\cal C}_{\rm b}(\tilde{\bf x})\setminus\{{\bf x}^*\}}} \Phi_{\rm u}({\cal C}_{{\rm a}_{\rm s}}({\bf x}))+\Phi_{\rm u}({\cal C}_{{\rm a}_{\rm s}}({\bf x}))}{W(1-\eta_{\rm a})}}-1\right).
\end{multline*}
From this step, the final expression of rate coverage for ORA can be derived on similar lines of the derivation for IRA. 
\end{IEEEproof}
We conclude this Section with the rate coverages of a single-tier macro-only network and a two-tier HetNet with fiber-backhauled SBSs which will be used for comparing the performances of IRA and ORA in Section~\ref{sec::results}.  The former can be obtained from \eqref{eq::rate::coverage::IRA} by setting $\lambda_{\rm s}=0$ and the later can be obtained from \eqref{eq::rate::ideal::backhaul} following the steps outlined in the proof of  Theorem~\ref{theorem::rate::coverage}.  
\begin{cor} \label{corr::rate::coverage::other::networks}For a single tier macro-only network, $\pr$ is given by:
\begin{equation}
\label{eq::rate::coverage::macro::only}
   \pr(\rho)  =\sum\limits_{n=1}^{\infty}\nbbP\bigg(
 \sinr_{\rm a}({\bf 0})>
 2^
 {
  \frac{\rho n}{W}
 }-1\bigg){\tt K}_t\left(n;\frac{\lambda_{\rm m}}{{\cal A}_{\rm m}},\lambda_{\rm u}\right).
\end{equation}
For a two-tier HetNet with fiber-backhauled SBSs, $\pr$ is given by:
\begin{multline}\label{eq::rate::coverage::perfect::backhaul}
   \pr(\rho)  = {\cal A}_{\rm m}\sum\limits_{n=1}^{\infty}\nbbP\bigg(
 \sinr_{\rm a}({\bf 0})>
 2^
 {
  \frac{\rho n}{W}
 }-1\bigg){\tt K}_t\left(n;\frac{\lambda_{\rm m}}{{\cal A}_{\rm m}},\lambda_{\rm u}\right)\\ + {\cal A}_{\rm s}\sum\limits_{n=1}^{\infty}\nbbP\bigg({\sinr}_{\rm a}({\bf 0})>
2^{\frac{\rho n}{W}}-1\bigg) {\tt K}_t\left(n;\frac{\lambda_{\rm s}}{{\cal A}_{\rm s}},\lambda_{\rm u}\right).
\end{multline}  
\end{cor}
\section{Results and Discussions}\label{sec::results}
\subsection{Verification of Accuracy}
\begin{figure}
\centering
\includegraphics[scale = 0.5]{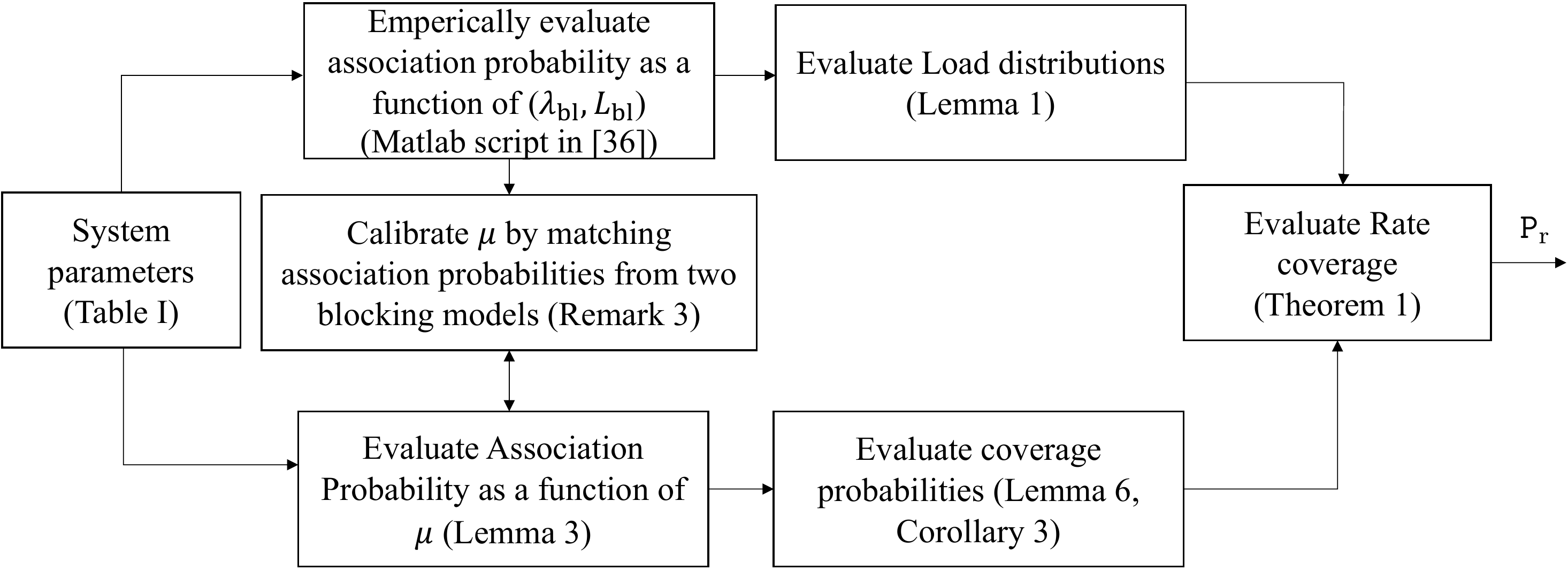}
\caption{Block diagram of the evaluation of rate coverage.}\label{fig::block::diagram}
\end{figure}

\begin{figure}
\centering
\subfigure[MBS coverage]{
 \includegraphics[scale=0.5]{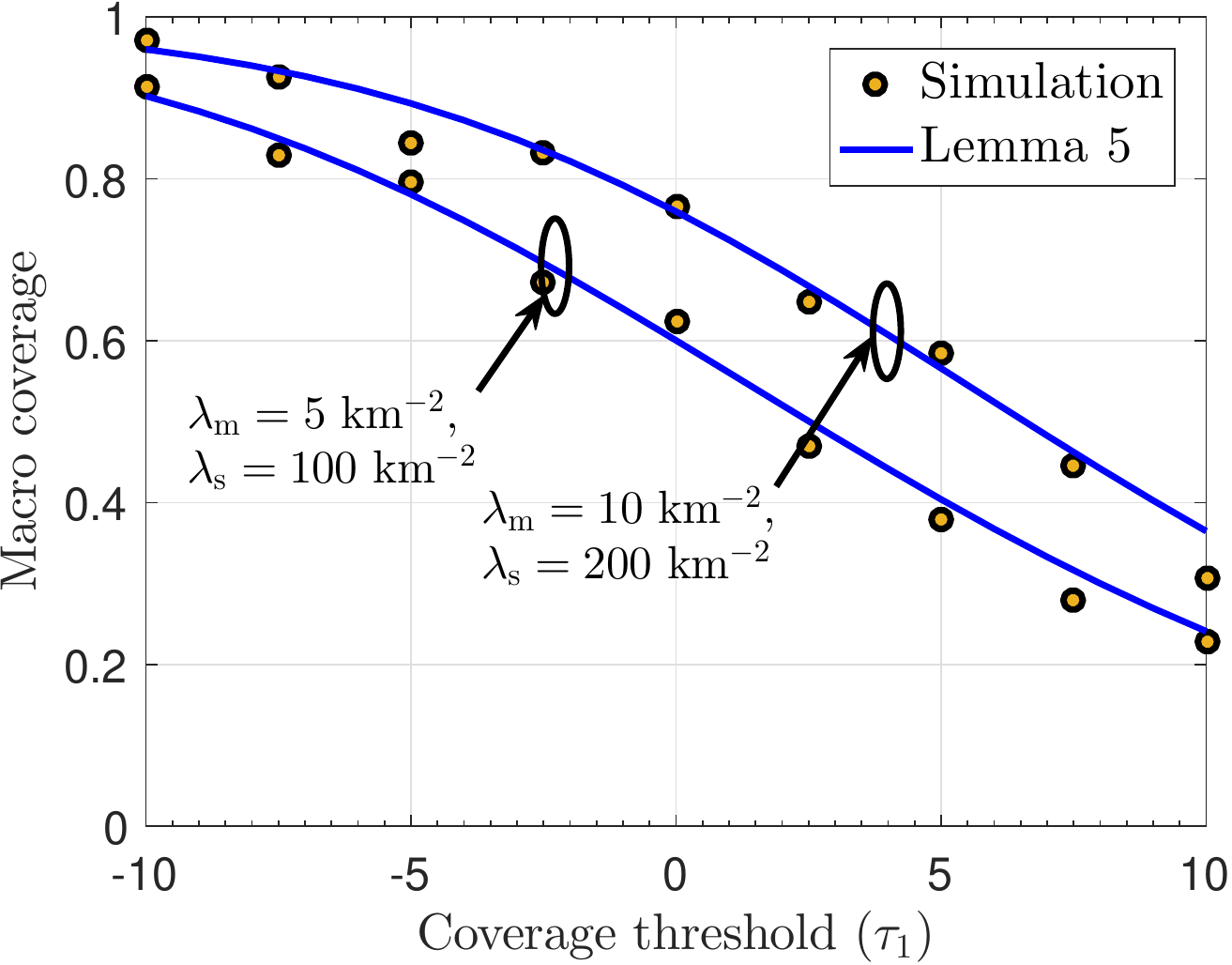}
            \label{fig::coverage::MBS}
        }\hfill
\subfigure[Joint SBS and backhaul coverage at $\tau_2$ = 5 dB]{
 \includegraphics[scale=0.5]{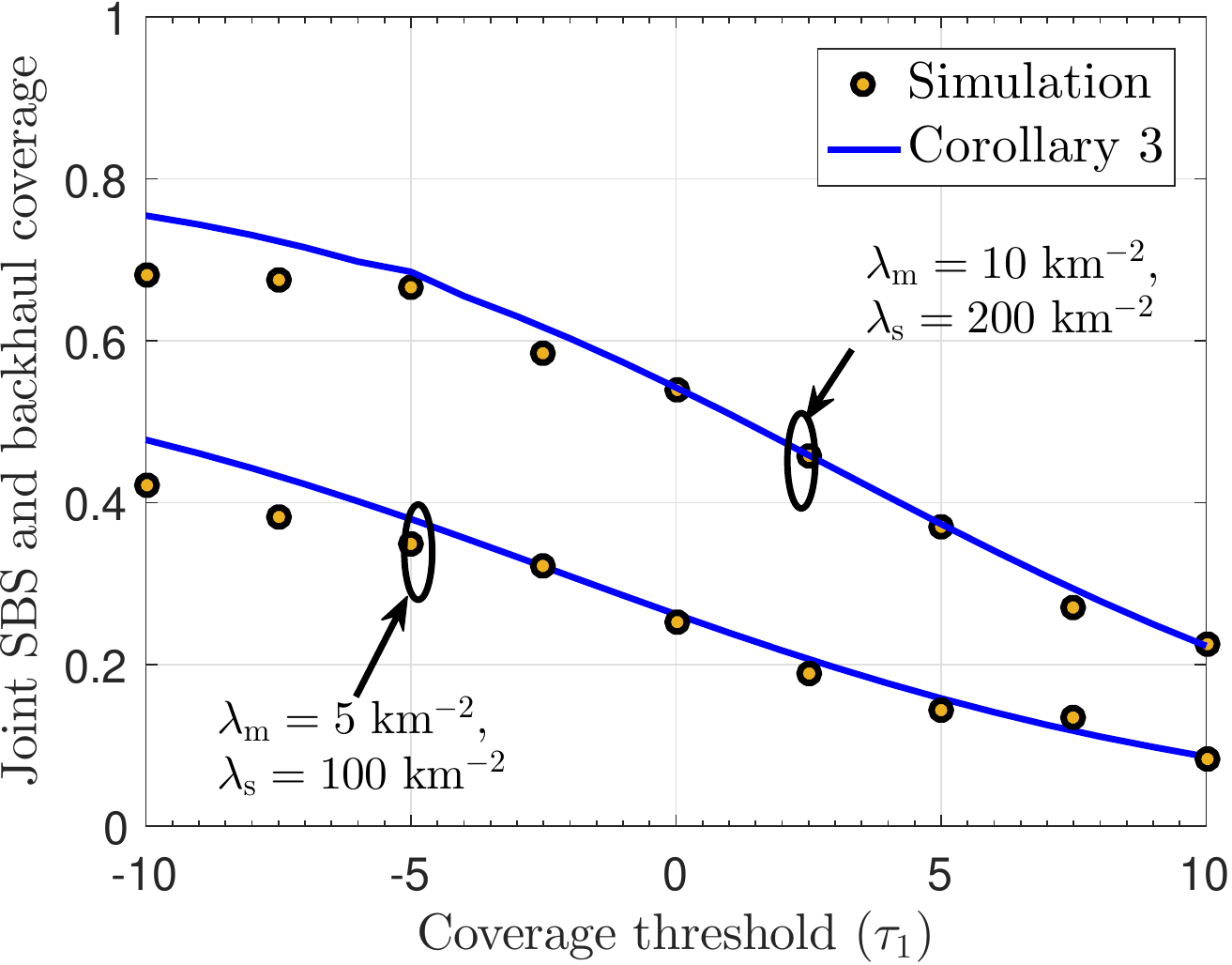}
           \label{fig::coverage::Joint}             
       }
 \caption{CCDFs of $\sinr$ distributions obtained from Monte Carlo simulation and analysis.}
             \label{fig::coverage}
\end{figure}

We now verify the accuracy of our analysis by comparing our analytical results with  Monte Carlo simulations of the network defined in Section~\ref{sec::sys::mod}. We further emphasize that the simulation of the network is a ``true'' simulation in the sense that it accounts for the spatial correlation of blocking while a lot of the existing works (including 3GPP) consider the independent blocking assumption in simulation. The values of the key system parameters are listed in Table~\ref{tab::parameters}. For each simulation, the number of iterations was set to $1\times 10^3$. The number of iterations is kept low because, as noted in Remark~\ref{rem::simulation}, the system-level simulation is extremely time consuming.  The primary reason is   the computation of every link state for which it is required to compute the intersection of each link with all the line segments of $\Phi_{\rm bl}$.  Following Remark~\ref{rem::fitting::mu}, we obtain $\mu = 200\ {\rm m}$ by  matching ${\cal A}_{\rm m}$ given by  \eqref{eq::association::probability::access} with its empirical value obtained by running the simulation scripts provided in~\cite{SahaIABCode}.  In Fig.~\ref{fig::coverage},   we plot the MBS and the joint SBS and backhaul coverages. The close match between theory and simulation validates our assumptions for the coverage analysis.   We now plot $\pr$ for IRA and ORA obtained from simulation and analysis (see Theorem~\ref{theorem::rate::coverage}). A block diagram of the evaluation of the analytical values of $\pr$ is provided in Fig.~\ref{fig::block::diagram}.
\begin{table}[t]
\centering
\caption{{Key system parameters and default values}}
\label{tab::parameters}\scalebox{0.8}
{
\begin{tabular}{|c|l|l|}
\hline
Notation & Parameter & Value \\ \hline
$P_{\rm m},\ P_{\rm s}$ & BS transmit powers  & 40, 20 dBm \\ \hline
$\alpha_{{k}_{{i},{\l}}}, \alpha_{{k}_{{i},{\rm n}}}$ ($\forall\ k\in\{{\rm a},{\rm b}\}$, $i\in\{{\rm m},{\rm s}\}$) & Path-loss exponent &  3.0,  4.0\\ \hline
$\beta_{{k}_{i}}$ ($\forall\ k\in\{{\rm a},{\rm b}\}$, $i\in\{{\rm m},{\rm s}\}$)& Path loss at 1 m & 70 dB \\\hline
$G_{\rm m}$, $G_{\rm s}$ & BS antenna main lobe gain & $18$ dB \\ \hline
$g_{\rm m}$, $g_{\rm s}$ & BS antenna side lobe gain & $-2$ dB \\ \hline
$G_{\rm u}$, $g_{\rm u}$ & UE antenna main and side lobe gains& $0$ dB \\ \hline
${\tt N}_0W$ & Noise power & \begin{tabular}[c]{@{}l@{}}$-174$ dBm/Hz+ $10\log_{10} W$ \\$+10$ dB {(noise-figure)}\end{tabular} \\ \hline
$\{\lambda_{\rm m},\lambda_{\rm s},\lambda_{\rm u}\}$& Density of MBS, SBS, and user PPP &$\{10,50, 1000\}$ km${}^{-2}$\\\hline
$T_{\rm m},T_{\rm s}$& Bias factors &1,1\\\hline 
$\lambda_{\rm u}$ & UE density & 1000 km${}^{-2}$\\\hline
$L_{\rm bl},\lambda_{\rm bl}$ & Blockage parameters & 5 m, 1500 km${}^{-2}$\\
\hline
\end{tabular}
}
\end{table}
For both strategies, we observe that $\pr$ obtained from our analysis closely follows $\pr$ obtained from simulation.   This further highlights the utility of our analytical expressions of $\pr$ which are considerably faster to evaluate than its computation by brute-force simulations.
\subsection{Optimal bandwidth partition for ORA}   
\begin{figure}[!htb]
\minipage{0.5\textwidth}
  \includegraphics[scale=0.5]{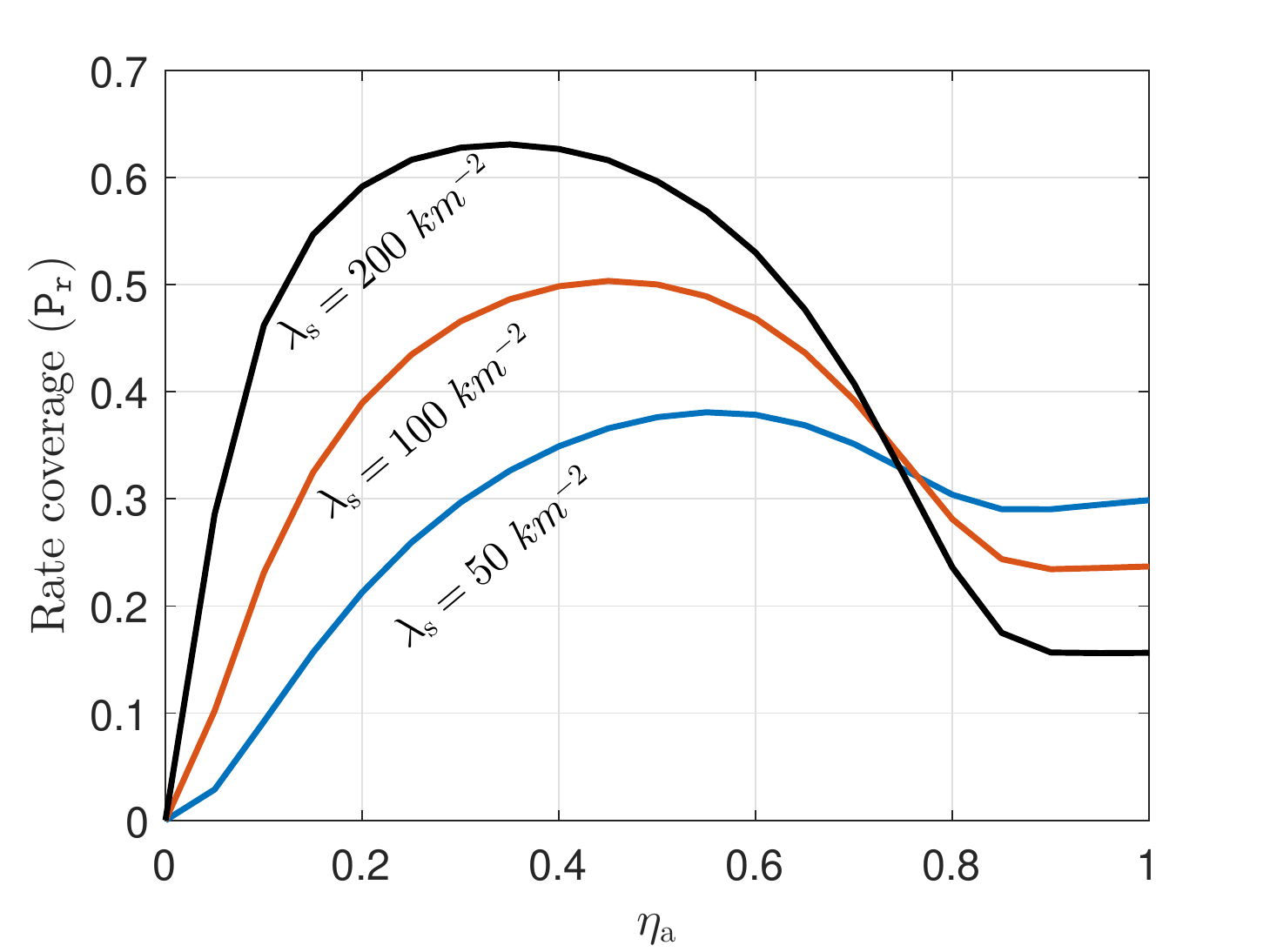}
          \caption{Rate coverage versus bandwidth partition factor for ORA ($\rho = 20$ Mbps).} \label{fig::vary::eta}
\endminipage\hspace{15pt}
\minipage{0.5\textwidth}
 \includegraphics[scale=0.5]{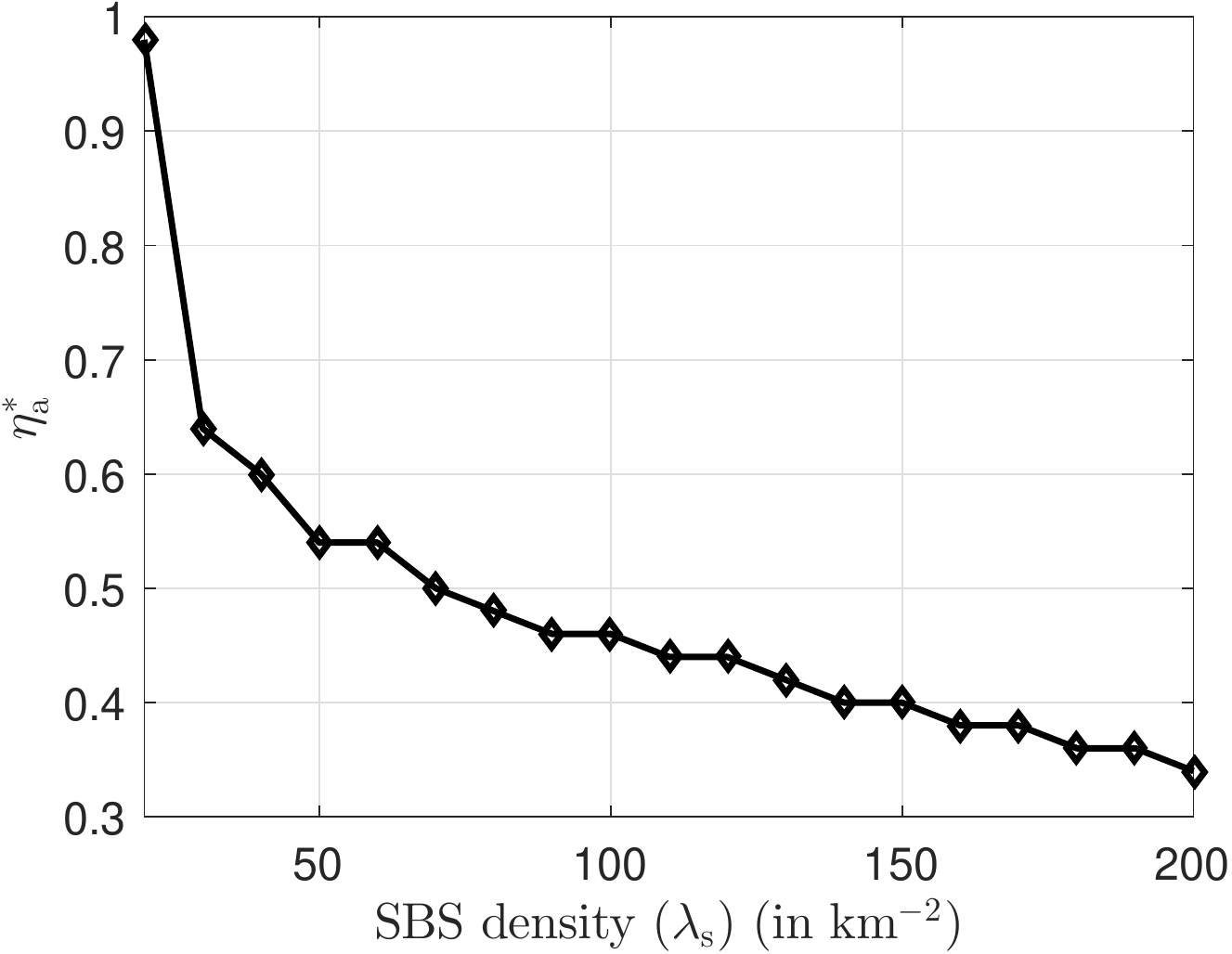}
        \caption{Optimum BW partition versus SBS density for ORA ($\rho = 20$ Mbps).}   \label{fig::etastar }
\endminipage
\end{figure}

In Fig.~\ref{fig::vary::eta}, we plot the variation of $\pr^{\rm ORA}$ with $\eta_{\rm a}$ for ORA. Note that $\eta_{\rm a}$ defines the BW split for ORA and  is hence a crucial system parameter~\cite{SahaBackhaul}.   While it is expected that $\pr^{\rm ORA}$ is quite sensitive to the choice of $\eta_{\rm a}$,  we observe that there is an optimal access-backhaul BW split ($\eta_{\rm a}^*$) for which $\pr^{\rm ORA}$ is maximized, i.e., $\eta_{\rm a}^* = \arg\max_{\eta}\pr^{\rm ORA}(\rho,\eta_{\rm a})$.  We also find that $\eta_{\rm a}^*$ decreases with increasing $\lambda_{\rm s}$ which is further evident from Fig.~\ref{fig::etastar }. This is because as $\lambda_{\rm s}$ increases, sufficient backhaul BW has to be reserved to support a given target data rate. Since this reduces the available access BW, it is clear that SBS densification provides diminishing returns for the overall rate performance of the network. We revisit this observation in Section~\ref{subsec::SBS::density}.
 \begin{figure}
\centering
\subfigure[IRA]{
 \includegraphics[scale=0.5]{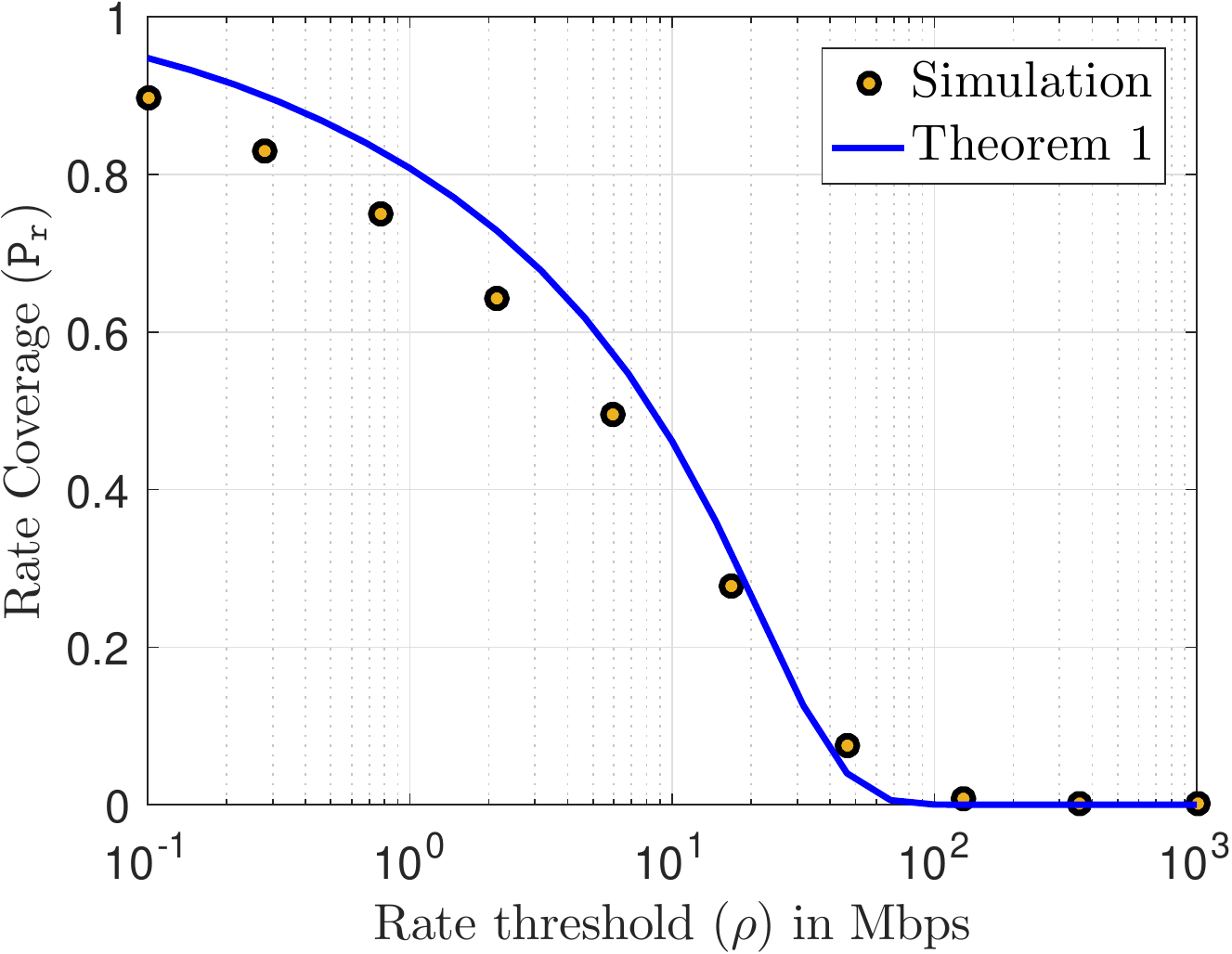}
            \label{fig::rate::coverage::IRA}
        }\hfill
\subfigure[ORA ($\eta_{\rm a}=0.8$)]{
 \includegraphics[scale=0.5]{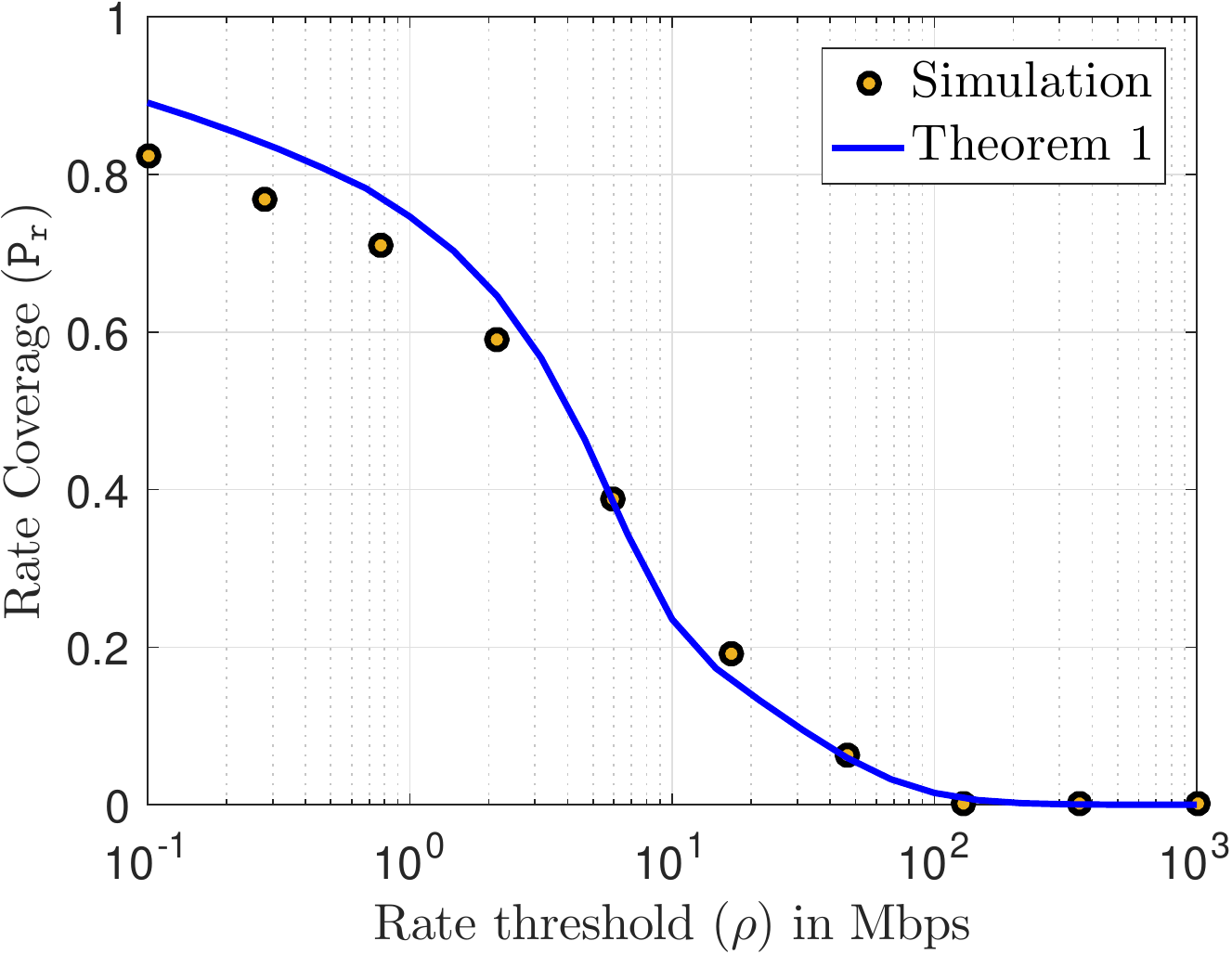}
           \label{fig::rate::coverage::ORA}             
       }
 \caption{Rate CCDF for IRA and ORA for the two-tier HetNet with IAB ($\lambda_{\rm m}=5$ km${}^{-2}$ and $\lambda_{\rm s}=100$ km${}^{-2}$).}
             \label{fig::rate::coverage}
\end{figure}
\subsection{User offloading and Rate Coverage}\label{subsec::bias}
\begin{figure}
\minipage{0.5\textwidth}
 \includegraphics[scale=0.5]{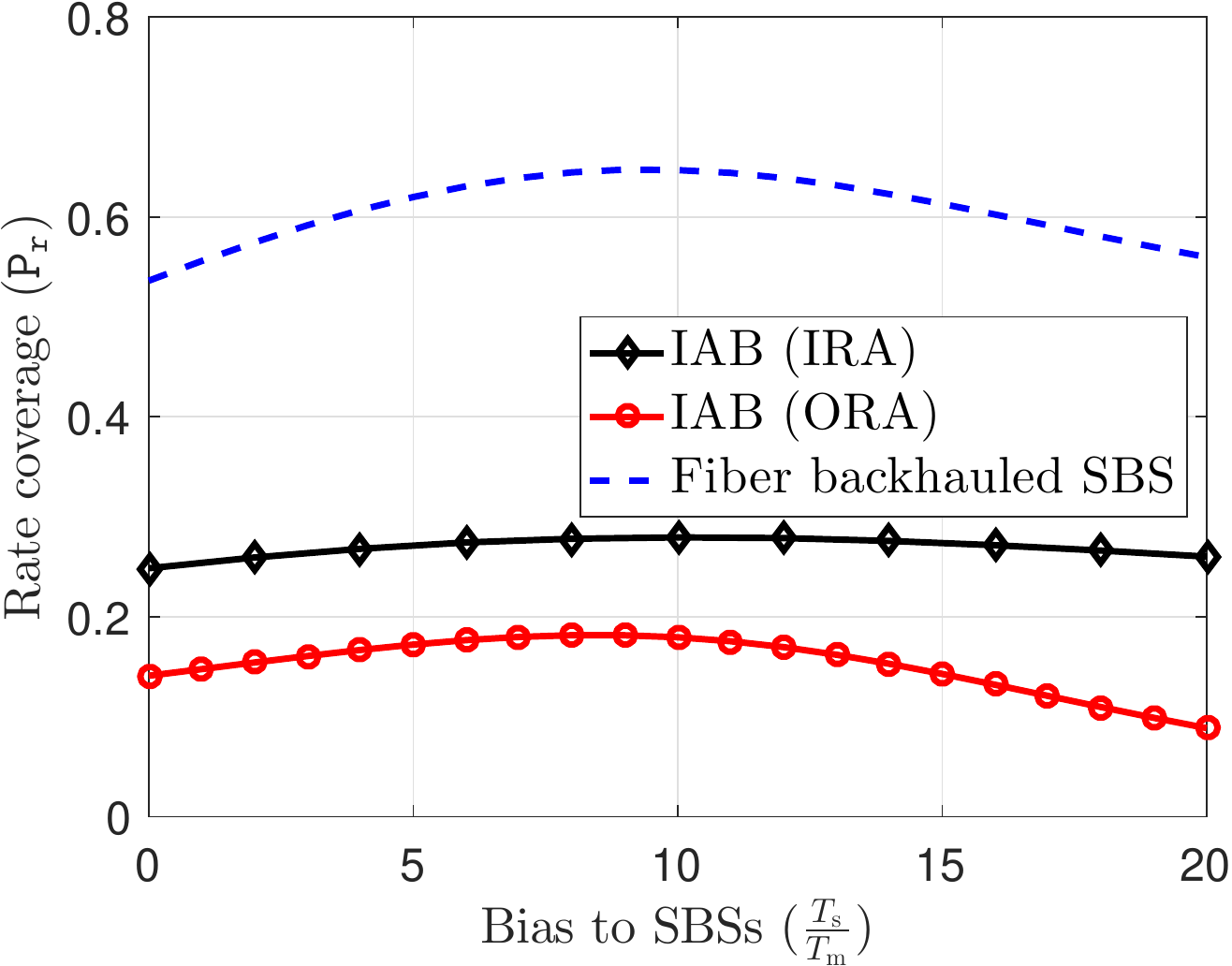}
        \caption{Rate coverage versus bias factor ($\rho = 50$ Mbps). 
}   \label{fig::vary::bias}
\endminipage\hspace{15pt}
\minipage{0.5\textwidth}
\includegraphics[scale=0.5]{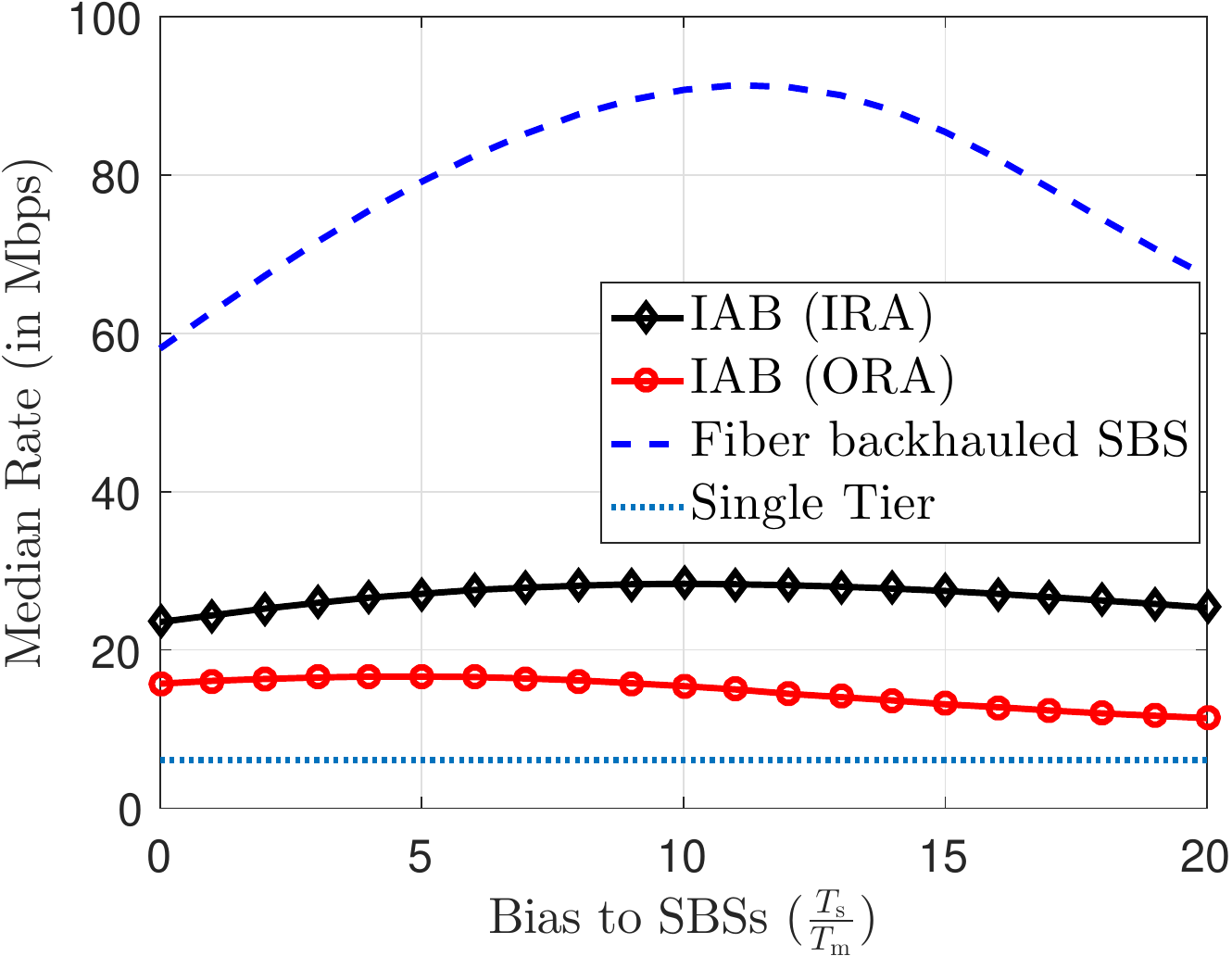}
\caption{Median rate versus  bias factor.} 
\label{fig::vary::bias::medianrate}
\endminipage
\end{figure}

We now offload more traffic to SBSs by increasing $T_{\rm s}$ ($T_{\rm m}=1$) and plot  $\pr$ and the median rate ($\rho_{50}$ where $\pr(\rho_{50})=0.5$) for IRA and ORA in Figs.~\ref{fig::vary::bias} and \ref{fig::vary::bias::medianrate}.  For comparison, we also plot $\pr$ and $\rho_{50}$ of a two-tier HetNet with fiber-backhauled SBSs (see Corollary~\ref{corr::rate::coverage::other::networks}). We observe that  $\pr$ and $\rho_{50}$ are  maximized at certain values of  $T_{\rm s}$.   
Also, $\pr^{\rm IRA}>\pr^{\rm ORA}$ and $\rho_{50}^{\rm IRA}>\rho_{50}^{\rm ORA}$, which are expected because of the system design of ORA, i.e. fixed $\eta_{\rm a}$ cannot cope up with the increase in backhaul load due to  increasing $T_{\rm s}$.   
What is interesting is that the improvement in $\pr$ and $\rho_{50}$ with $T_{\rm s}$ is much less prominent for IAB than $\pr$ for the HetNet with fiber-backhauled SBSs.  This is because  the offloaded UEs from $\Phi_{\rm m}$ to $\Phi_{\rm s}$ are not completely disappearing from the MBS load, i.e., they are coming back to the MBS load in the form of   increased backhaul load. Note that this phenomena is quite unique to the IAB design and does not occur for the HetNet with fiber-backhauled SBSs.   Thus, traffic offloading in  IAB-enabled HetNets is not as effective as in  HetNets with fiber-backhauled SBSs. However, as indicated by Fig.~\ref{fig::medianrate}, the two-tier network with IAB  still performs better than a single-tier macro-only network.
\subsection{SBS density and Rate coverage}\label{subsec::SBS::density}
\begin{figure}
\minipage{0.5\textwidth}
 \includegraphics[scale=0.5]{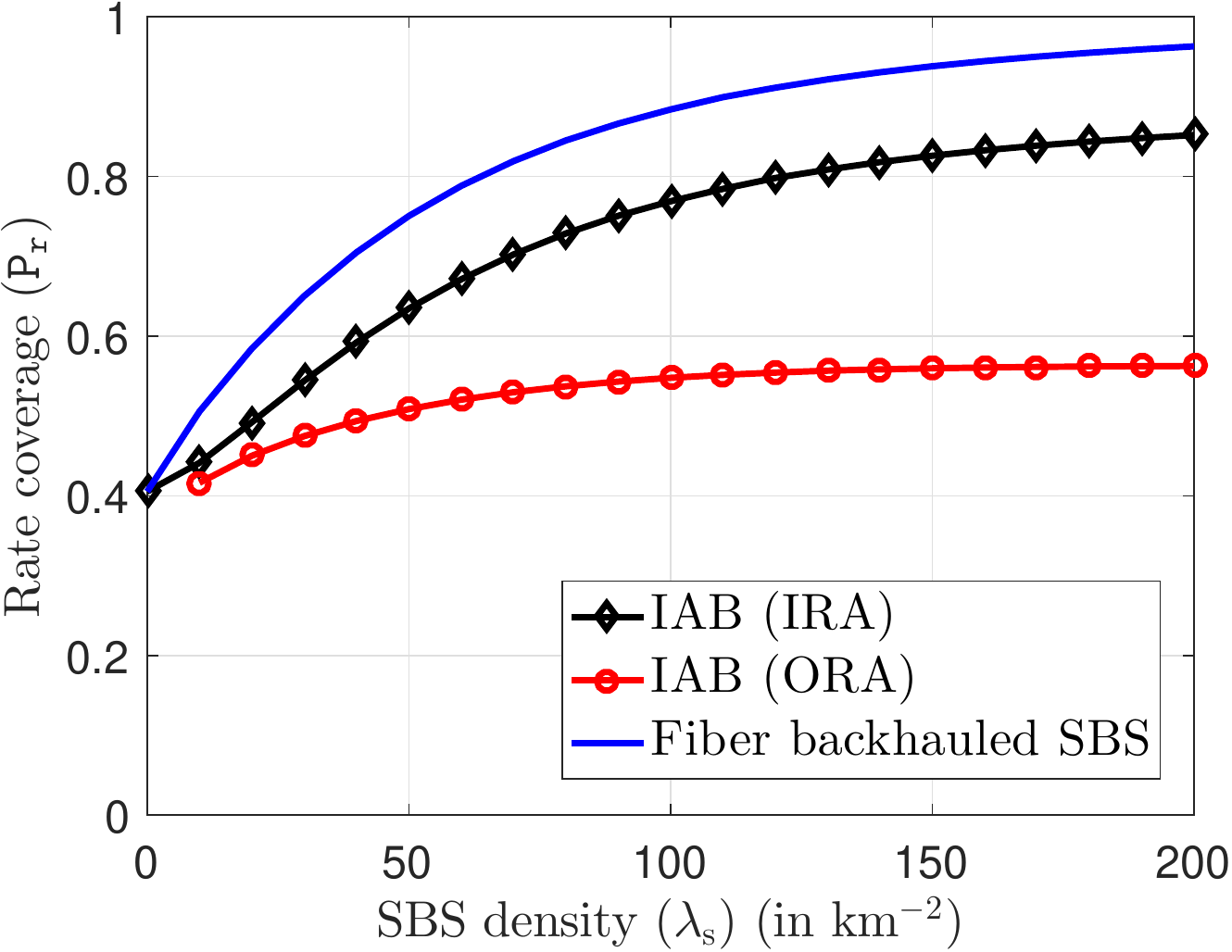}
\caption{Rate coverage versus SBS density ($\rho= 10$ Mbps, $\frac{T_{\rm s}}{T_{\rm s}}=10$ dB, $\eta_{\rm a}=0.8$).}\label{fig::vary::lambda_s}
\endminipage\hspace{15pt}
\minipage{0.5\textwidth}
\includegraphics[scale=0.5]{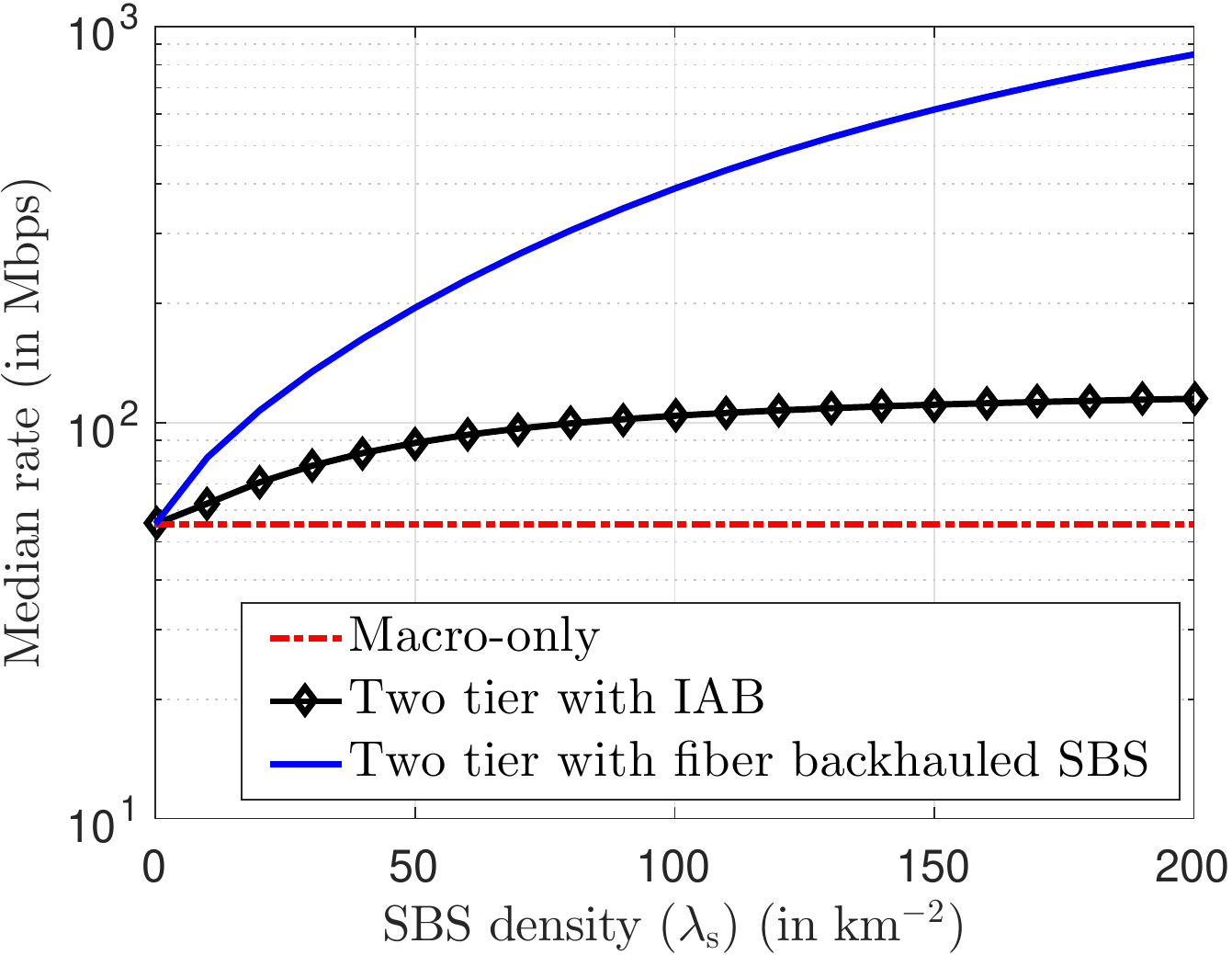}
\caption{Median rate versus SBS density ($\frac{T_{\rm s}}{T_{\rm s}}=10$ dB).}\label{fig::medianrate}
\endminipage
\end{figure}

We plot the the variation of $\pr$ with $\lambda_{\rm s}$ in Fig.~\ref{fig::vary::lambda_s}.  
As expected, $\pr$ increases with $\lambda_{\rm s}$. However, while $\pr$ steadily increases with $\lambda_{\rm s}$ for the fiber-backhauled SBSs, $\pr$ tends to  saturate for IAB. This effect is more prominent in Fig.~\ref{fig::medianrate}, where we plot $\rho_{50}$ versus $\lambda_{\rm s}$. 
  Figs.~\ref{fig::vary::lambda_s} and \ref{fig::medianrate} clearly illustrate the realistic gain of SBS densification in HetNets. Although the two-tier HetNet is  prominently advantageous over a single tier macro-only network, the assumption of fiber backhaul for all SBSs  leads to an overestimation of the rate improvement of HetNets with increasing $\lambda_{\rm s}$. Since  the overall rate is limited by the rate on the backhaul link, increasing $\lambda_{\rm s}$ decreases the rate supported by the wireless  backhual  as the BW is shared by more number of SBSs. 
\section{Conclusion}
In this paper, we proposed a tractable model of an IAB-enabled mm-wave two-tier HetNet  where all MBSs have access to fiber backhaul and the SBSs are wirelessly backhauled by the MBSs. For this network, we derived the CCDF of  downlink end-user data rate assuming that the total BW at the MBS is split between access and backhaul links by dynamic or static partitions. While the blockages in mm-wave communication and  the two hop links from MBS to UE over SBS due to the IAB setup impose analytical challenges for the exact characterization of the rate distributions, we propose reasonable approximations  that allow us to obtain easy-to-compute expressions of rate coverage.  Using these expressions, we obtain some useful system  insights of the multi-tier IAB design such as the impact of traffic offloading and SBS density on data rate. 

This work can be extended in multiple directions. While the SBSs in our  IAB setup are not very different from the LTE layer 3 relays~\cite{polese2018distributed}, one can extend this setup for studying multi-hop backhaul where the packets reach from the MBSs to UEs through multiple SBSs. This model can also  be used to build  a SG-based framework   for analyzing transmission delays in an IAB-enabled network. Further,  one can formulate a cost model (as done in \cite{suryaprakash2014analysis}) to study the benefits and possible revenues of deploying wireless backhaul over traditional fiber backhaul.
\appendix
\subsection{Proof of Lemma~\ref{lemm::joint::coverage}}
\label{app::lemm::joint::coverage}
The MBS coverage can be written as $\nbbP({\tt SINR}_{\rm a}({\bf 0})>\tau|{\bf x}^*\in\Phi_{\rm m}) = 
$
\begin{align*}
& 
\nbbP\left(\frac{P_{\rm m}\beta_{{\rm a}_{\rm m}}G_{\rm m} G_{\rm u} h_{0,{\bf x}^*}L_{{\rm a}_{\rm m}}(\|{\bf x}^*\|)^{-1}}{I_{{\rm a}_{\rm s}}+{\tt N}_0W}>\tau\bigg|{\bf x}^*\in\Phi_{\rm m}\right)=\nbbP\left(h_{0,{\bf x}^*}>\tau\frac{I_{{\rm a}_{\rm s}}+{\tt N}_0W}{P_{\rm m}\beta_{{\rm a}_{\rm m}}G_{\rm m}G_{\rm u}{{L_{{\rm a}}^*}}^{-1}}\bigg|{\bf x}^*\in\Phi_{\rm m}\right)\\&
=\nbbE\left[e^{-\tau\frac{I_{\rm s}+{\tt N}_0W}{P_{\rm m}\beta_{{\rm a}_{\rm m}}G_{\rm m}G_{\rm u}{L_{{\rm a}}^*}^{-1}}}\bigg|{\bf x}^*\in\Phi_{\rm m}\right]
=\nbbE\left[\nbbE\left[e^{-\frac{\tau L^*_{{\rm a}}}{P_{\rm m}\beta_{{\rm a}_{\rm m}}G_{\rm m}G_{\rm u}}I_{{\rm a}_{\rm s}}}\right]e^{-\frac{\tau {\tt N}_0W L_{{\rm a}}^*}{P_{\rm m}\beta_{{\rm a}_{\rm m}}G_{\rm m}G_{\rm u}}}\bigg|{\bf x}^*\in\Phi_{\rm m}\right].
\end{align*}
The first step follows from  Assumption~\ref{assumption::sinr::snr}-a. 
In the last step, the expectations inside the product are conditional expectations given $L_{\rm a}^*$ while the outer expectation is with respect to $L_{\rm a}^*$. 
Now focusing on the first term of the product, 
\scalebox{0.9}{$\nbbE\big[e^{-\frac{\tau L_{{\rm a}}^*}{P_{\rm m}\beta_{{\rm a}_{\rm m}}G_{\rm m}G_{\rm u}}I_{{\rm a}_{\rm s}}}\big]=\nbbE\bigg[\exp\bigg({-\frac{\tau L_{\rm a}^*}{P_{\rm m}\beta_{{\rm a}_{\rm m}}G_{\rm m}G_{\rm u}}\sum\limits_{\substack{{z}\in{\cal L}_{{\rm a}_{\rm s}},\\z>\Omega_{{\rm s},{\rm m}}L_{\rm a}^*}}P_{\rm s} h_{0,{\bf x}} \beta_{{\rm a}_{\rm s}} {\psi}_{{\rm a}_{\rm s}}  z^{-1}
}\bigg)\bigg]$}
\begin{multline*}\scalebox{0.9}{$
\stackrel{(a)}{=}\nbbE\bigg[
\prod\limits_{\substack{{z}\in{\cal L}_{{\rm a}_{\rm s}},\\z>\Omega_{{\rm s},{\rm m}}L_{\rm a}^*}}\nbbE\left[e^{-\frac{\tau P_{\rm s}h_{0,{\bf x}}\beta_{{\rm a}_{\rm s}}{\psi}_{{\rm a}_{\rm s}} L_{{\rm a}}^*}{P_{\rm m}\beta_{{\rm a}_{\rm m}}G_{\rm m}G_{\rm u}z}}\right]\bigg|{L}_{\rm a}^*\bigg]=\nbbE\bigg[\prod\limits_{\substack{{z}\in{\cal L}_{{\rm a}_{\rm s}},\\z>\Omega_{{\rm s},{\rm m}}{L_{\rm a}}^*}}\nbbE\bigg[\frac{1}{1+\frac{\tau P_{\rm s}\beta_{{\rm a}_{\rm s}}{\psi}_{{\rm a}_{\rm s}} L_{{\rm a}}^*}{P_{\rm m}\beta_{{\rm a}_{\rm m}}G_{\rm m}G_{\rm u}z}}\bigg]\bigg|L_{\rm a}^*\bigg]$}\\\scalebox{0.9}{$=\prod\limits_{{\cal G}\in{\cal M}_{{\rm a}_{\rm s}}}
\exp\bigg({-\int\limits_{\Omega_{{\rm s},{\rm m}}L_{{\rm a}}^*}^{\infty}\bigg(1-\frac{1}{1+\frac{\tau P_{\rm s}\beta_{{\rm a}_{\rm s}}{\cal G} L_{{\rm a}}^*}{P_{\rm m}\beta_{{\rm a}_{\rm m}}G_{\rm m} G_{\rm u}z}}\bigg)p_{\cal G}\lambda_{{\rm a}_{{\rm s}}}(z)\:{\rm d}z}\bigg).$}
\end{multline*}
Step $(a)$ follows from the assumption that $\{h_{0,{\bf x}}\}$ is an i.i.d. sequence of exponential random variables.  The last step follows from the fact that conditioned on ${L_{\rm a}}^*$, the pathloss process of the BSs of $\Phi_{\rm s}$ with effective antenna gain ${\cal G}$ is a thinned version of the PPP ${\cal L}_{{\rm a}_{\rm s}}\cap [0,\Lambda_{{\rm s},{\rm m}}L_{{\rm a}}^*]^c$ with thinning probability  $p_{\cal G}$~\cite{Bai_mmWave}. Hence we apply  the probability generating functional of PPP~\cite{baccelli2009stochastic} to compute the product over the point process.     
The final expression in \eqref{eq::sinr::macro::coverage}  is obtained by deconditioning over the distribution of $L_{{\rm a}}^*$ whose PDF is given by \eqref{eq::serving::dist::distribution}. 
 Now, the conditional joint SBS and backhaul coverage  can be expressed as:  $\nbbP({\sinr}_{\rm a}({\bf 0})>\tau_1,\snr_{\rm b}({\bf x}^*)>\tau_2|{\bf x}^*\in\Phi_{\rm s})=\scalebox{0.95}{ $\sum_{s({\bf x}^*) \in \{\l,\n\}} \nbbP({\sinr}_{\rm a}({\bf 0})>\tau_1,\snr_{\rm b}({\bf x}^*)>\tau_2|{\bf x}^*\in\Phi_{\rm s},s({\bf x}^*,{\bf 0}))\nbbP(s({\bf x}^*,{\bf 0})|{\bf x}^*\in \Phi_{\rm s})=$}$
\begin{align*}
&\scalebox{0.95}{$ = \sum\limits_{s({\bf x}^*) \in \{\l,\n\}} \nbbP({\sinr}_{\rm a}({\bf 0})>\tau_1,\snr_{\rm b}({\bf x}^*)>\tau_2|{\bf x}^*,{\bf x}^*\in\Phi_{\rm s},s({\bf x}^*,{\bf 0}))\frac{\nbbP(s({\bf x}^*,{\bf 0}),{\bf x}^*\in \Phi_{\rm s})}{\nbbP({\bf x}^*\in \Phi_{\rm s})}.$}
\end{align*}
The second term is equal to ${{\cal A}_{{\rm s}_k}}/{{\cal A}_{\rm s}}$ and   $\nbbP({\sinr}_{\rm a}({\bf 0})>\tau_1,\snr_{\rm b}({\bf x}^*)>\tau_2|{\bf x}^*\in\Phi_{\rm s},s({\bf x}^*,{\bf 0}))=$
\begin{multline*}
\scalebox{0.95}{$\nbbP\left(\frac{P_{\rm s}\beta_{{\rm a}_{\rm s}}G_{\rm s}G_{\rm u} h_{0,{\bf x}^*}{L_{{\rm a}}^*}^{-1}}{I_{{\rm a}_{\rm s}}+{\tt N}_0W}>\tau_1,\frac{P_{\rm m}\beta_{{\rm a}_{\rm m}}G_{\rm m}G_{\rm s} h_{{\bf x}^*,\tilde{\bf x}}{\tilde{L}_{{\rm b}}}^{-1}}{{\tt N}_0W}>\tau_2\bigg|{\bf x}^*\in\Phi_{\rm s},s({\bf x}^*,{\bf 0})\right)$}\\
\scalebox{0.95}{$=\nbbP\left(h_{0,{\bf x}^*}>\frac{\tau_1(I_{{\rm a}_{\rm s}}+{\tt N}_0W){L_{{\rm a}}^*}}{P_{\rm s}\beta_{{\rm a}_{\rm s}}G_{\rm s}G_{\rm u} },h_{{\bf x}^*,\tilde{\bf x}}>\frac{\tau_2{\tt N}_0W\tilde{L}_{{\rm b}}}{P_{\rm m}\beta_{{\rm a}_{\rm m}}G_{\rm m}G_{\rm s}}\bigg|{\bf x}^*\in\Phi_{\rm s},s({\bf x}^*,{\bf 0})\right)$}
\\\scalebox{0.95}{$=\nbbE\left[e^{-\frac{\tau_1(I_{{\rm a}_{\rm s}}+{\tt N}_0W)L_{{\rm a}}^*}{P_{\rm s}\beta_{{\rm a}_{\rm s}}G_{\rm s}G_{\rm u} }}\nbbE\left[e^{-\frac{\tau_2{\tt N}_0W\tilde{L}_{{\rm b}}}{P_{\rm m}\beta_{{\rm a}_{\rm m}}G_{\rm m}G_{\rm s}}}\bigg|L_{\rm a}^*,s({\bf x}^*,{\bf 0}),{\bf x}\in\Phi_{\rm s}\right]\bigg|{\bf x}^*\in\Phi_{\rm s},s({\bf x}^*,{\bf 0})\right].$}
\end{multline*}
The first step follows from Assumption~\ref{assumption::sinr::snr}-a.  Here $\tilde{L}_{\rm b}=\min({\cal L}_{{\rm b}_{\rm m}}|{\bf x}\in \Phi_{\rm s},s({\bf x},{\bf 0}))$. Note that the outer expectation is with respect to $L_{\rm a}^*|{\bf x}^*\in\Phi_{\rm s},s({\bf x}^*,{\bf 0})$ whose PDF is given by  Corollary~\ref{cor::pathloss::special::case::association::servingPDF}. The first exponential term  can be handled exactly as the MBS coverage. The inner expectation is with respect to  $\tilde{L}_{\rm b}|{\bf x}^*\in \Phi_{\rm s},s({\bf x}^*,{\bf 0}),L_{\rm a}^*$ whose PDF is given by:  
$f_{\tilde{L}_{\rm b}}(l|{\bf x}^*\in \Phi_{\rm s},s({\bf x}^*,{\bf 0})= t ,L_{\rm a}^*) = \nbbE_{\theta^*}[\tilde{\lambda}_{{\rm b}_t}(l;L_{\rm a}^*,\theta^*)e^{-\tilde{\Lambda}_{{\rm b}_t}((0,l];L_{\rm a}^*,\theta^*)}] =\tilde{\lambda}_{{\rm b}_t}(l;L_{\rm a}^*,0)e^{-\tilde{\Lambda}_{{\rm b}_t}((0,l];L_{\rm a}^*,0)} , \ l>0,  
$ 
where $\tilde{\lambda}_{{\rm b}_t}(l;L_{\rm a}^*,\theta^*)$ and $\tilde{\Lambda}_{{\rm b}_t}((0,l];L_{\rm a}^*,\theta^*)$ are obtained from    Lemma~\ref{lemm::joint::pathloss::distribution}.
 Note that the expectation with respect to $\theta^*$ (which is a uniform random variable within $(0,2\pi]$) can be simplified since it can be shown that the function under the exception is invariant to $\theta^*$. 

\bibliography{LoabBalancingHetNetv13.bbl}

\begin{thebibliography}{10}
\providecommand{\url}[1]{#1}
\csname url@samestyle\endcsname
\providecommand{\newblock}{\relax}
\providecommand{\bibinfo}[2]{#2}
\providecommand{\BIBentrySTDinterwordspacing}{\spaceskip=0pt\relax}
\providecommand{\BIBentryALTinterwordstretchfactor}{4}
\providecommand{\BIBentryALTinterwordspacing}{\spaceskip=\fontdimen2\font plus
\BIBentryALTinterwordstretchfactor\fontdimen3\font minus
  \fontdimen4\font\relax}
\providecommand{\BIBforeignlanguage}[2]{{%
\expandafter\ifx\csname l@#1\endcsname\relax
\typeout{** WARNING: IEEEtran.bst: No hyphenation pattern has been}%
\typeout{** loaded for the language `#1'. Using the pattern for}%
\typeout{** the default language instead.}%
\else
\language=\csname l@#1\endcsname
\fi
#2}}
\providecommand{\BIBdecl}{\relax}
\BIBdecl

\bibitem{thompson20145g}
J.~Thompson, X.~Ge, H.-C. Wu, R.~Irmer, H.~Jiang, G.~Fettweis, and S.~Alamouti,
  ``5{G} wireless communication systems: Prospects and challenges,'' \emph{IEEE
  Commun. Mag}, vol.~52, no.~2, pp. 62--64, 2014.

\bibitem{SahaBackhaul}
C.~Saha, M.~Afshang, and H.~S. Dhillon, ``Bandwidth partitioning and downlink
  analysis in millimeter wave integrated access and backhaul for 5{G},''
  \emph{IEEE Trans. on Wireless Commun.}, vol.~17, no.~12, pp. 8195--8210, Dec.
  2018.

\bibitem{saha2017integrated}
------, ``Integrated mm{W}ave access and backhaul in 5{G}: Bandwidth
  partitioning and downlink analysis,'' in \emph{Proc., IEEE Int. Conf. Commun.
  (ICC)}, May 2018.

\bibitem{dhillon2012modeling}
H.~S. Dhillon, R.~K. Ganti, F.~Baccelli, and J.~G. Andrews, ``Modeling and
  analysis of {$K$}-tier downlink heterogeneous cellular networks,'' \emph{IEEE
  Journal on Sel. Areas in Commun.}, vol.~30, no.~3, pp. 550--560, Apr. 2012.

\bibitem{OffloadingSingh}
S.~Singh, H.~S. Dhillon, and J.~G. Andrews, ``Offloading in heterogeneous
  networks: Modeling, analysis, and design insights,'' \emph{IEEE Trans. on
  Wireless Commun.}, vol.~12, no.~5, pp. 2484--2497, May. 2013.

\bibitem{accessbackhaul3gpp}
{3GPP TR 38.874}, ``{NR}; {S}tudy on integrated access and backhaul,'' Tech.
  Rep., 2017.

\bibitem{polese2018distributed}
M.~Polese, M.~Giordani, A.~Roy, D.~Castor, and M.~Zorzi, ``Distributed path
  selection strategies for integrated access and backhaul at mm{W}aves,'' 2018,
  available online: arxiv.org/abs/1805.04351.

\bibitem{m2018maxmin}
M.~N. Kulkarni, A.~Ghosh, and J.~G. Andrews, ``Max-min rates in self-backhauled
  millimeter wave cellular networks,'' 2018, available online:
  arxiv.org/abs/1805.01040.

\bibitem{Delayaware}
J.~Garc{\'\i}a-Rois, R.~Banirazi, F.~J. Gonz{\'a}lez-Casta{\~n}o, B.~Lorenzo,
  and J.~C. Burguillo, ``Delay-aware optimization framework for proportional
  flow delay differentiation in millimeter-wave backhaul cellular networks,''
  \emph{IEEE Trans. on Commun.}, vol.~66, no.~5, pp. 2037--2051, May 2018.

\bibitem{polese2018end}
M.~Polese, M.~Giordani, A.~Roy, S.~Goyal, D.~Castor, and M.~Zorzi, ``End-to-end
  simulation of integrated access and backhaul at mm{W}aves,'' in \emph{2018
  IEEE 23rd International Workshop on Computer Aided Modeling and Design of
  Communication Links and Networks (CAMAD)}, Sep. 2018, pp. 1--7.

\bibitem{DeRenzoBackhaul}
A.~Mesodiakaki, F.~Adelantado, L.~Alonso, M.~D. Renzo, and C.~Verikoukis,
  ``Energy- and spectrum-efficient user association in millimeter-wave backhaul
  small-cell networks,'' \emph{IEEE Trans. on Vehicular Technology}, vol.~66,
  no.~2, pp. 1810--1821, Feb. 2017.

\bibitem{Madhusudhanan2014}
P.~Madhusudhanan, J.~G. Restrepo, Y.~Liu, T.~X. Brown, and K.~R. Baker,
  ``Downlink performance analysis for a generalized shotgun cellular system,''
  \emph{IEEE Trans. on Wireless Commun.}, vol.~13, no.~12, pp. 6684--6696, Dec.
  2014.

\bibitem{DiRenzoUplink2016}
M.~D. Renzo and P.~Guan, ``Stochastic geometry modeling and system-level
  analysis of uplink heterogeneous cellular networks with multi-antenna base
  stations,'' \emph{IEEE Trans. on Commun.}, vol.~64, no.~6, pp. 2453--2476,
  Jun. 2016.

\bibitem{Singh_association_cell}
S.~Singh, F.~Baccelli, and J.~G. Andrews, ``On association cells in random
  heterogeneous networks,'' \emph{IEEE Wireless Commun. Letters}, vol.~3,
  no.~1, pp. 70--73, Feb. 2014.

\bibitem{Rate6658810}
H.~S. Dhillon and J.~G. Andrews, ``Downlink rate distribution in heterogeneous
  cellular networks under generalized cell selection,'' \emph{IEEE Wireless
  Commun. Letters}, vol.~3, no.~1, pp. 42--45, Feb. 2014.

\bibitem{suryaprakash2014analysis}
V.~Suryaprakash and G.~P. Fettweis, ``An analysis of backhaul costs of radio
  access networks using stochastic geometry,'' in \emph{Proc. IEEE Int. Conf.
  Commun. (ICC)}, Jun. 2014, pp. 1035--1041.

\bibitem{SinghResourcePartition}
S.~Singh and J.~G. Andrews, ``Joint resource partitioning and offloading in
  heterogeneous cellular networks,'' \emph{IEEE Trans. on Wireless Commun.},
  vol.~13, no.~2, pp. 888--901, Feb. 2014.

\bibitem{DiRenzoBackhaulHyperdense}
F.~J. Martin-Vega, M.~D. Renzo, M.~C. Aguayo-Torres, G.~Gomez, and T.~Q. Duong,
  ``Stochastic geometry modeling and analysis of backhaul-constrained
  hyper-dense heterogeneous cellular networks,'' in \emph{17th International
  Conference on Transparent Optical Networks (ICTON)}, Jul. 2015, pp. 1--4.

\bibitem{Ganti-self-backhaul}
A.~Sharma, R.~K. Ganti, and J.~K. Milleth, ``Joint backhaul-access analysis of
  full duplex self-backhauling heterogeneous networks,'' \emph{IEEE Trans. on
  Wireless Commun.}, vol.~16, no.~3, pp. 1727--1740, Mar. 2017.

\bibitem{tabassum2016analysis}
H.~Tabassum, A.~H. Sakr, and E.~Hossain, ``Analysis of massive {MIMO}-enabled
  downlink wireless backhauling for full-duplex small cells,'' \emph{IEEE
  Trans. on Commun.}, vol.~64, no.~6, pp. 2354--2369, Jun. 2016.

\bibitem{QuekBackhaul}
G.~Zhang, T.~Q. Quek, M.~Kountouris, A.~Huang, and H.~Shan, ``Fundamentals of
  heterogeneous backhaul design--{A}nalysis and optimization,'' \emph{IEEE
  Trans. on Commun.}, vol.~64, no.~2, pp. 876--889, Feb 2016.

\bibitem{DhillonCaire2015}
H.~S. Dhillon and G.~Caire, ``Wireless backhaul networks: Capacity bound,
  scalability analysis and design guidelines,'' \emph{IEEE Trans. on Wireless
  Commun.}, vol.~14, no.~11, pp. 6043--6056, Nov. 2015.

\bibitem{Bai_mmWave}
T.~Bai, R.~Vaze, and R.~W. Heath, ``Analysis of blockage effects on urban
  cellular networks,'' \emph{IEEE Trans. on Wireless Commun.}, vol.~13, no.~9,
  pp. 5070--5083, Sep. 2014.

\bibitem{direnzo2015stochastic}
M.~Di~Renzo, ``Stochastic geometry modeling and analysis of multi-tier
  millimeter wave cellular networks,'' \emph{IEEE Trans. on Wireless Commun.},
  vol.~14, no.~9, pp. 5038--5057, Sep. 2015.

\bibitem{mmWaveHetNetTurgut}
E.~Turgut and M.~C. Gursoy, ``Coverage in heterogeneous downlink millimeter
  wave cellular networks,'' \emph{IEEE Trans. on Commun.}, vol.~65, no.~10, pp.
  4463--4477, Oct. 2017.

\bibitem{SinghKulkarniSelfBackhaul}
S.~Singh, M.~N. Kulkarni, A.~Ghosh, and J.~G. Andrews, ``Tractable model for
  rate in self-backhauled millimeter wave cellular networks,'' \emph{IEEE
  Journal on Sel. Areas in Commun.}, vol.~33, no.~10, pp. 2196--2211, Oct.
  2015.

\bibitem{AndrewsMMWave}
J.~G. Andrews, T.~Bai, M.~N. Kulkarni, A.~Alkhateeb, A.~K. Gupta, and R.~W.
  Heath, ``Modeling and analyzing millimeter wave cellular systems,''
  \emph{IEEE Trans. Commun.}, vol.~65, no.~1, pp. 403--430, Jan. 2017.

\bibitem{Kulkarni_backhaul_asilomar}
S.~Singh, M.~N. Kulkarni, and J.~G. Andrews, ``A tractable model for rate in
  noise limited mm{W}ave cellular networks,'' in \emph{Proc. IEEE Asilomar},
  Nov. 2014, pp. 1911--1915.

\bibitem{Elshaer_mmwave_association}
H.~Elshaer, M.~N. Kulkarni, F.~Boccardi, J.~G. Andrews, and M.~Dohler,
  ``Downlink and uplink cell association with traditional macrocells and
  millimeter wave small cells,'' \emph{IEEE Trans. on Wireless Commun.},
  vol.~15, no.~9, pp. 6244--6258, Sep. 2016.

\bibitem{3GPPNR}
{3GPP TR 38.901}, ``Study on channel model for frequencies from 0.5 to 100
  {GH}z,'' Tech. Rep., 2017.

\bibitem{Aditya_blocking_letter}
S.~Aditya, H.~S. Dhillon, A.~F. Molisch, and H.~Behairy, ``Asymptotic
  blind-spot analysis of localization networks under correlated blocking using
  a {P}oisson line process,'' \emph{IEEE Wireless Commun. Letters}, vol.~6,
  no.~5, pp. 654--657, Oct. 2017.

\bibitem{Aditya_Blindspot}
S.~Aditya, H.~S. Dhillon, A.~F. Molisch, and H.~M. Behairy, ``A tractable
  analysis of the blind spot probability in localization networks under
  correlated blocking,'' \emph{IEEE Trans. on Wireless Commun.}, vol.~17,
  no.~12, pp. 8150--8164, Dec. 2018.

\bibitem{loadbalancingAndrews2014}
J.~G. Andrews, S.~Singh, Q.~Ye, X.~Lin, and H.~S. Dhillon, ``An overview of
  load balancing in {HetNets}: old myths and open problems,'' \emph{IEEE
  Wireless Commun.}, vol.~21, no.~2, pp. 18--25, Apr. 2014.

\bibitem{DownlinkUplinkCellAssociationmmWave}
H.~Elshaer, M.~N. Kulkarni, F.~Boccardi, J.~G. Andrews, and M.~Dohler,
  ``Downlink and uplink cell association with traditional macrocells and
  millimeter wave small cells,'' \emph{IEEE Trans. on Wireless Commun.},
  vol.~15, no.~9, pp. 6244--6258, Sep. 2016.

\bibitem{SinghJointRate2015}
S.~Singh, X.~Zhang, and J.~G. Andrews, ``Joint rate and {SINR} coverage
  analysis for decoupled uplink-downlink biased cell associations in hetnets,''
  \emph{IEEE Trans. on Wireless Commun.}, vol.~14, no.~10, pp. 5360--5373, Oct.
  2015.

\bibitem{SahaIABCode}
C.~Saha and H.~S. Dhillon, ``Matlab code for the computation of association
  probability in 5{G} mm-wave {H}et{N}et with integrated access and backhaul,''
  2019, available at:
  {github.com/stochastic-geometry/Load-balancing-5G-mmwave}.

\bibitem{baccelli2009stochastic}
F.~Baccelli and B.~Blaszczyszyn, \emph{Stochastic Geometry and Wireless
  networks{,} Volume 1- Theory}.\hskip 1em plus 0.5em minus 0.4em\relax NOW:
  Foundations and Trends in Networking, 2009.

\bibitem{krishnan2016spatio}
S.~Krishnan and H.~S. Dhillon, ``Spatio-temporal interference correlation and
  joint coverage in cellular networks,'' \emph{IEEE Trans. on Wireless
  Commun.}, vol.~16, no.~9, pp. 5659--5672, Sep. 2017.

\end{thebibliography}
\end{document}